\documentclass[a4paper,fleqn, usenatbib]{mnras}

\usepackage{newtxtext,newtxmath}

\usepackage{graphicx}
\usepackage{array}
\usepackage{amssymb}
\usepackage{natbib}
\usepackage{float}
\usepackage{color}
\usepackage[utf8]{inputenc}  
\usepackage{amsmath}  
\usepackage{comment}
\usepackage{footnote}
\usepackage{longtable}
\usepackage{caption}
\usepackage{supertabular}
\usepackage{threeparttable}
\usepackage[figuresright]{rotating}

\newcommand{\degree}{\ensuremath{^\circ}}
\bibpunct{(}{)}{;}{a}{,}{,}

 \title{Polarization of seven MBM clouds at high galactic latitude}

\author[Neha et al.]{Neha, S.$^{1,2}$\thanks{Email: pathakneha.sharma@gmail.com; neha.astro18@gmail.com},
Maheswar, G.$^{1,3}$,
Soam, A.$^{4}$,
Chang Won Lee$^{4,5}$
\\\\
$^{1}$Aryabhatta Research Institute of Observational Sciences (ARIES), Nainital (Uttarakhad) $-$ 263002, India.\\
$^{2}$Pt. Ravishankar Shukla University, Raipur (Chhattisgarh) $-$ 492010, India.\\
$^{3}$Indian Institute of Astrophysics, Block II, Koramangala, Bangalore, 560 034, India\\
$^{4}$Korea Astronomy $\&$ Space Science Institute (KASI), 776 Daedeokdae-ro, Yuseong-gu, Daejeon, Republic of Korea.\\
$^{5}$University of Science Technology, 217 Gajungro, Yuseong-gu, 305-333 Daejeon, Republic of Korea.
}
\date{Accepted 2018, February 16. Received 2018, February 15; in original form 2018, January 04}

\pubyear{2015}

\begin{document}
\label{firstpage}
\pagerange{\pageref{firstpage}--\pageref{lastpage}}
\maketitle

\begin{abstract}
We made R-band polarization measurements of 234 stars towards the direction of MBM 33-39 cloud complex. The distance of MBM 33-39 complex was determined as $120\pm10$ pc using polarization results and near-infrared photometry from the 2MASS survey. The magnetic field geometry of the individual cloud inferred from our polarimetric results reveals that the field lines are in general consistent with the global magnetic field geometry of the region obtained from the previous studies. This implies that the clouds in the complex are permeated by the interstellar magnetic field. Multi-wavelength polarization measurements of a few stars projected on the complex suggest that the size of the dust grains in these clouds is similar to those found in the normal interstellar medium of the Milky Way. We studied a possible formation scenario of MBM 33-39 complex by combining the polarization results from our study and from the literature and by identifying the distribution of ionized, atomic and molecular (dust) components of material in the region.

\end{abstract}




\section{Introduction}

How cold, molecular and dense structures, that are sites of star formation, are formed from hot, atomic and diffuse gas in the interstellar medium is still unclear. Stellar feedback processes like the expansion of HII regions \citep{1980ApJ...239..173B, 1995ApJ...441..702V, 1995ApJ...455..536P} and supernova blast waves \citep{1987ApJ...317..190M, 1999ApJ...518..748G, 2000MNRAS.315..479D, 2005A&A...436..585D, 2011ApJ...731...13N} are considered to be the plausible mechanisms by which hot, over pressurized bubbles are created. As these shells expand, they sweep up the surrounding medium and inject it into high galactic latitudes. The gas then tends to slide down back guided by the magnetic field lines creating large-scale flows and gather in location where infalling streams converge and shock leading to the formation of high-latitude molecular clouds \citep{2000A&A...359.1124H, 2007ApJ...658L..99I, 2008ApJ...687..303I, 2008ApJ...680..336S, 2009ApJ...695..248H, 2012ApJ...759...35I, 2013ASPC..476..115H, 2016ApJ...833...10I}. The high-latitude clouds (HLCs) generally do not have any internal energy source, but most HLCs may be affected by the mean interstellar far-ultraviolet (FUV) radiations \citep{1988ApJ...334..771V}. The relatively lower densities and proximity to the Sun make the HLCs ideal objects to study the penetration of FUV radiations, which may have an important role in the formation and evolution of these clouds and the role of the magnetic field could also be crucial in the formation process. Most of the HLCs are not gravitationally bound and hence, they do not show any signature of star formation, but a few HLCs, e.g., MBM 12 \citep{2001ApJ...560..287L} and MBM 20 \citep[L1642;][]{2014A&A...563A.125M} show evidence of star formation. Thus, the star-formation is not considered to be a dominant process in most HLCs \citep{1996ApJS..106..447M}.

\begin{center}
\begin{table*}
\caption{Parameters of high-Galactic latitude clouds studied in this work.}\label{HLC_para}
\begin{minipage}{\textwidth}
\begin{tabular}{c c c c c c c}
\hline
S. No.	&Cloud			&$l$		&$b$		&Opacity$\dagger$	&Distance$\ddagger$&Other\\
		&Identification	&($\degree$)&($\degree$)&Class		&(pc)&Designations\\
\hline
1.		&MBM 33			&359.06		&36.75		&4/5		&$89^{+18}_{-21}$&LDN 1778A, LDN 1778B, LDN 1780\\
2.		&MBM 34			&2.36		&35.67		&-			&$110^{+27}_{-34}$&\\
3.		&MBM 35			&6.56		&38.13		&-			&$89^{+17}_{-25}$&\\
4.		&MBM 36			&4.18		&35.75		&5			&$105^{+7}_{-7}$&LDN 134, LDN 134A\\
5.		&MBM 37			&5.70		&36.62		&5			&$121^{+10}_{-16}$&LDN 183, LDN 184\\
6.		&MBM 38			&8.22		&36.62		&-			&$77^{+24}_{-24}$&CB 68\\
7.		&MBM 39			&11.39		&36.22		&-			&$94^{+15}_{-11}$&\\
\hline
\end{tabular}
\end{minipage}
$\dagger$ Opacity classes have been taken from \cite{2002A&A...383..631D}.\\
$\ddagger$ Distances are taken from \citet{2014ApJ...786...29S}.
\end{table*}
\end{center}

\citet{2002A&A...383..631D}, from a heterogeneous set of surveys based on infrared, CO and optical detection methods \citep{1985ApJ...295..402M, 1986ApJ...304..466K, 1988ApJ...334..815D, 1996ApJS..106..447M, 1998ApJ...507..507R}, have compiled a total of 439 HLCs ($b\geq30\degree$). Molecular clouds are classified based on their visual extinction (A$_{V}$) along the line of sight through the clouds \citep{1988ApJ...334..771V}. Clouds with A$_{V}<1$ magnitude are termed as diffuse clouds and those with A$_{V}>5$ magnitudes are called the dark clouds. The third category of clouds, termed as translucent clouds, are those having values of A$_{V}$ lying in the range from $1-5$ magnitudes. Out of 439, 393 clouds are found to have central $A_{V}$ values less than 1 magnitude and are therefore classified as diffuse clouds. The remaining 45 clouds are classified as translucent clouds. LDN 1457 (MBM 12) is the only HLC, which was classified with $A_{V}>5$ \citep{2008hsf2.book..813M}. However, \cite{2011PASJ...63S...1D} estimated A$_{V}$ values for dust clouds using Two Micron All Sky Survey Point-Source Catalog \citep[2MASS PSC;][]{2003yCat.2246....0C, 2006AJ....131.1163S} and found A$_{V}>5$ for three MBM clouds, e.g., 5.44 $\pm$ 0.15 for MBM 20, 8.56 $\pm$ 0.15 for MBM 36 and 6.76 $\pm$ 0.16 for MBM 37. Thus, HLCs  are predominately diffuse or translucent types. The above classification scheme also reflects the kind of astrochemistry taking place in these clouds \citep{1988ApJ...334..771V}. While the chemistry in diffuse clouds is mostly due to photo-processes, dark clouds are influenced by collisional processes. Thus, it may be argued that the translucent clouds represent an intermediate stage between the diffuse and the dark clouds making them the ideal sources to study the initial stages of cloud formation. 


Many authors studied the dynamical condensation process of the ISM in shocks \citep{2000ApJ...532..980K, 2002Ap&SS.281...67I, 2005AIPC..784..318I} and in converging flows \citep{1999A&A...351..309H, 2005A&A...433....1A, 2005ApJ...633L.113H, 2007A&A...465..431H, 2006ApJ...643..245V}. Based on magnetohydrodynamics simulations, it was shown that molecular clouds can be created behind a shock wave if it is moving roughly parallel to the mean magnetic field lines \citep{2007ApJ...658L..99I, 2008ApJ...687..303I, 2012ApJ...759...35I, 2016ApJ...833...10I}. A perpendicular magnetic field will decrease the compression of the post-shock gas and thus suppresses the process necessary for the rapid flow fragmentations and build-up of high density regions \citep{2000A&A...359.1124H, 2004ApJ...616..288B, 2007ApJ...658L..99I, 2008ApJ...674..316H, 2008ApJ...683..786H}.

In this paper, we present results of our multi-wavelength ($V$, $R$, $I$) polarimetric observations of an HLC complex, MBM 33-39. The parameters of the seven clouds, studied in this work, are listed in Table \ref{HLC_para}. The distance of the clouds which ranges from $77$ pc to $121$ pc \citep{1989A&A...223..313F, 1992BaltA...1..163C, 1996ApJS..106..447M, 2003A&A...411..447L, 2014ApJ...786...29S} shows that these are local clouds lying at a galactic height ranging from $\sim$45 pc to $\sim$70 pc. The local scale height of the molecular gas layer is estimated to be $\sim$70 pc with an uncertainty of 25\% \citep{1991IAUS..144...41B}. Therefore, this complex could be considered as an example of a group located just at the boundary of the molecular gas layer in the solar vicinity. The main objectives of our study are to map the magnetic field geometry of the individual clouds in the complex and understand the relationships between the field geometry and their density structures. In this paper, section \ref{sec:obs} explains the observations and the procedures of data reduction. Our results and discussion are described in section \ref{sec:res_dis}. At last, we finalize our paper by summarizing the results in section \ref{sec:conclude}.

\section{Observations and Data Reduction}\label{sec:obs}

Polarimetric observations were performed using the Aries IMaging POLarimeter \citep[AIMPOL,][]{2004BASI...32..159R}. This polarimeter is used as back-end instrument at Cassegrain focus of the 104-cm Sampurnanand telescope, which is located at Aryabhatta Research Institute of Observational Sciences (ARIES), Nainital, India. It is coupled with TK 1024$\times$1024 $pixel^{2}$ CCD camera and a half-wave plate (HWP) modulator and a Wollaston prism beam-splitter are used in between. The Wollaston prism splits the light rays into two images; conventionally knows as ordinary and extraordinary. The observations were made in $\lambda_{Rkc_{eff}}$= 0.760 $\mu$m photometric band. We also observed a few bright stars towards MBM 33, MBM 34 and MBM 35 in multi-wavelengths (V, R \& I). The plate scale of CCD is 1.48 arcsec$/$pixel and field of view is $\sim$8 arcmin in diameter. The full width at half maximum (FWHM) varies from 2 to 3 pixels. The Read out noise and gain of CCD are 7.0 $e^{-1}$  and 11.98 $e^{-1}$/ADU, respectively. Table \ref{tab:obslog} represents the details of the observations.

\begin{table}
\centering
\caption{Log of observations in the R$_{kc}$ filter ($\lambda_{Rkc_{eff}}$= 0.760 $\mu$m).}\label{tab:obslog}
\begin{tabular}{c c}\hline
 Cloud Name        &  Date of observations (year, month,date)\\
\hline
 MBM 33           & 2015, February, 21; 2015, March 18;      \\        
 & 2015, April, 22  \\
 				  & 2015, May, 14, 18, 21, 22, 24 \\
 MBM 34           & 2013, April, 19 \\
 				  & 2013, May, 3, 4, 13, 14, 15, 16	\\
 MBM 35           & 2016, April, 6 \\
 MBM 36           & 2016, May, 5 \\
 MBM 38           & 2016, May, 10 \\
 MBM 39           & 2016, May, 11 \\
\hline
\end{tabular}
\end{table}

We executed standard aperture photometry using the Image Reduction \& Analysis Facility (IRAF) software to extract the fluxes of ordinary ({\it $I_{o}$}) and extraordinary ({\it $I_{e}$}) images for all the observed stars with a good signal-to-noise ratio. The ratio {\it {R($\alpha$)}} is given by
\begin{equation}
R(\alpha) = \frac{\frac{{I_{e}}(\alpha)}{{I_{o}}(\alpha)}-1} {\frac{I_{e}(\alpha)} {I_{o}(\alpha)}+1} =  P cos(2\theta - 4\alpha)
\end{equation}
where {\it P} is the degree of polarization; $\theta$ is the polarization position angle and {\it $\alpha$} is the position of the fast axis of HWP at $0\degree$, $22.5\degree$, $45\degree$ and $67.5\degree$ corresponding to four normalized Stokes parameters, q[R($0\degree$)], u[R($22.5\degree$)], $q_{1}$[R($45\degree$)] and $u_{1}$[R($67.5\degree$)], respectively. We estimate the uncertainties in normalized Stokes parameters ($\sigma_R$)($\alpha$)($\sigma_q$, $\sigma_u$, $\sigma_{q1}$, $\sigma_{u1}$) in percent using the following equation \citep{1998A&AS..128..369R},
\begin{equation}
\sigma_R(\alpha)= \frac{\sqrt{N_{e}+N_{o}+2N_{b}}}{N_{e}+N_{o}}
\end{equation}
where $N_{o}$ and $N_{e}$ are the counts of ordinary and extraordinary images, respectively, and $N_{b}$[= ({$N_{be}$}$+${$N_{bo}$})/2] is the mean background counts around the extraordinary and ordinary images. 

\citet{2004BASI...32..159R, 2008MNRAS.388..105M, 2011MNRAS.411.1418E} observed un-polarized standard stars regularly and found that the instrumental polarization of AIMPOL on the 104-cm Sampurnanand Telescope is approximately invariable. The instrumental polarization is found to be less than $\sim$0.1\%. To determine the reference direction of the polarizer, we have observed two polarized standard stars HD 154445 and HD 155197, adopted from \citet{1992AJ....104.1563S}, on each observing night. Additionally, another two standard stars HD 19820 and HD 25443 were also observed in one night. The standard stars observed by us to calibrate our observations are reported in Table \ref{tab:std}.

\begin{table}
\caption{Polarized standard stars observed in R$_{kc}$ band.}\label{tab:std}
\begin{tabular}{lll}\hline
Date of     &P $\pm$ $\epsilon_P$ 	&  $\theta$ $\pm$ $\epsilon_{\theta}$  \\
Obs.		&(\%)            		& ($\degree$)                           \\\hline
\multicolumn{3}{c}{{\bf HD 19820}} \\
\multicolumn{3}{c}{($^\dagger$Standard values: 4.526 $\pm$ 0.025\%, 114.46 $\pm$ 0.16$\degree$)}\\
21 Feb 2015 & 4.4 $\pm$ 0.1     & 113 $\pm$ 2 \\\hline

\multicolumn{3}{c}{{\bf HD 25443}}\\
\multicolumn{3}{c}{($^\dagger$Standard values: 4.734 $\pm$ 0.045\%, 133.65 $\pm$ 0.28$\degree$)}\\
21 Feb 2015 & 4.7 $\pm$ 0.1     & 134 $\pm$ 2 \\\hline

\multicolumn{3}{c}{{\bf HD 154445}}\\ 
\multicolumn{3}{c}{($^\dagger$Standard values: 3.683 $\pm$ 0.072\%, 88.92 $\pm$ 0.56$\degree$)}\\
18 Mar 2015 & 3.5 $\pm$ 0.1     & 89 $\pm$ 2 \\
22 Apr 2015 & 3.5 $\pm$ 0.2     & 90 $\pm$ 1 \\
14 May 2015 & 3.4 $\pm$ 0.2     & 89 $\pm$ 1 \\
18 May 2015 & 3.4 $\pm$ 0.2     & 89 $\pm$ 1 \\
21 May 2015 & 3.4 $\pm$ 0.2     & 90 $\pm$ 1 \\
22 May 2015 & 3.8 $\pm$ 0.2     & 88 $\pm$ 2 \\
24 May 2015 & 3.5 $\pm$ 0.2     & 87 $\pm$ 2 \\
06 Apr 2016 & 3.7 $\pm$ 0.1     & 89 $\pm$ 1 \\
05 May 2016 & 3.5 $\pm$ 0.1     & 89 $\pm$ 1 \\
10 May 2016 & 3.4 $\pm$ 0.1     & 88 $\pm$ 1 \\
\hline

\multicolumn{3}{c}{{\bf HD 155197}}\\
\multicolumn{3}{c}{($^\dagger$Standard values: 4.274 $\pm$ 0.027\%, 102.88 $\pm$ 0.18$\degree$)}\\
18 Mar 2015 & 4.2 $\pm$ 0.1     & 103 $\pm$ 2 \\
22 Apr 2015 & 4.0 $\pm$ 0.3     & 103 $\pm$ 1 \\
14 May 2015 & 4.1 $\pm$ 0.1     & 103 $\pm$ 1 \\
18 May 2015 & 4.1 $\pm$ 0.1     & 102 $\pm$ 1 \\
21 May 2015 & 4.1 $\pm$ 0.1     & 103 $\pm$ 1 \\
22 May 2015 & 4.3 $\pm$ 0.1     & 103 $\pm$ 1 \\
24 May 2015 & 4.1 $\pm$ 0.1     & 102 $\pm$ 1 \\
06 Apr 2016 & 4.0 $\pm$ 0.1     & 102 $\pm$ 1 \\
05 May 2016 & 3.7 $\pm$ 0.1     & 103 $\pm$ 1 \\
10 May 2016 & 4.3 $\pm$ 0.1     & 103 $\pm$ 1 \\
11 May 2016 & 4.0 $\pm$ 0.1     & 102 $\pm$ 1 \\
\hline
\end{tabular}

$\dagger$  Values in R band from \citet{1992AJ....104.1563S} \\
\end{table}

\section{RESULTS}\label{sec:res_dis}

\begin{figure}
\centering
\resizebox{0.48\textwidth}{14.5cm}{\includegraphics{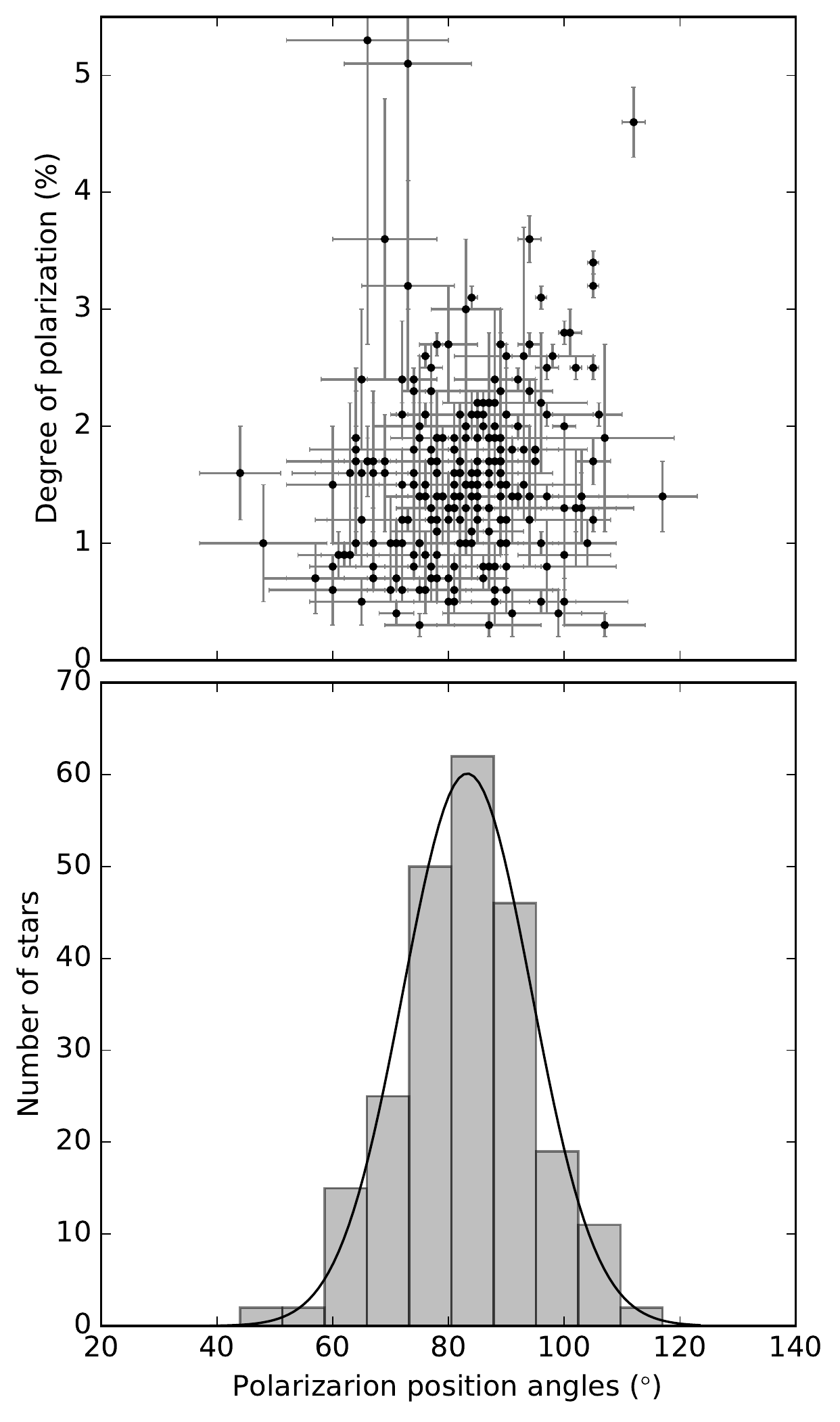}}
\caption{The degree of polarization vs. the polarization position angle of 234 stars observed towards MBM 33-39 is shown in the upper panel. The histogram of the $\theta_{P}$ along with a Gaussian fit is shown in the lower panel.}\label{fig:allhist}
\end{figure}

Polarization measurements of 234 stars were obtained towards the direction of seven MBM clouds. The average values of polarization measurements are presented in Table \ref{pol_res} (the detailed results are shown in Table \ref{ch7:Polresult}). The columns 1 and 2 give the cloud name and the number of stars observed towards individual clouds, respectively. In columns 3 and 4 we give the mean values and the standard deviations of P and $\theta_{P}$, respectively. In Figure \ref{fig:allhist}, we present the P\% vs. $\theta_{P}$ in the upper panel and the histogram of the $\theta_{P}$ in the lower panel. A Gaussian fit to the histogram is also shown. The mean value and the standard deviation for the 234 stars are 1.6 $\pm$ 0.8\% and 83 $\pm$ 12$\degree$, respectively.

\begin{table}
\begin{center}
\caption{Average values of polarization measurements.}\label{pol_res}
\begin{tabular}{c c c c c c c}
\hline
Cloud	&Number of stars	&$<P>$	&$<\theta_{P}>$\\	
Name	&observed			&(\%)	&($\degree$)	\\
\hline
MBM 33	&49&$1.3\pm0.6$&$79\pm15$\\
MBM 34	&59&$1.6\pm0.4$&$83\pm9$\\
MBM 35	&21&$1.5\pm0.3$&$84\pm7$\\
MBM 36	&20&$2.1\pm0.5$&$77\pm9$\\
MBM 37	&26&$2.6\pm0.6$&$94\pm8$\\
MBM 38	&33&$1.1\pm0.4$&$87\pm11$\\
MBM 39	&26&$0.8\pm0.5$&$78\pm13$\\
\hline
\end{tabular}\\
\end{center}
Note: Column 1 represents the name of the molecular cloud; column 2 shows the total number of stars observed towards the individual cloud; column 3 denotes the mean values and standard deviations of the degree of polarization, and column 4 denotes the mean values and standard deviations of the polarization position angle.
\end{table}

\section{DISCUSSION}

\subsection{Distance of the complex}

\subsubsection{From polarization measurements}
The distance estimated using parallax measurements and the polarization measurements of stars projected onto the clouds could be considered as more reliable because of the fact that both the measurements are independent of the properties of the observed stars. In addition to this, the measured polarization position angles provide an additional clue to the presence of the most dominant dust layer along a given line of sight. While the foreground stars would show relatively large dispersion in the values of polarization position angles, the stars that lie behind the cloud layer, inferred from the high degree of polarization values, would show less dispersion with a constant value depending upon the plane of the sky orientation of the local magnetic field. 

\begin{figure}
\resizebox{8.5cm}{12cm}{\includegraphics{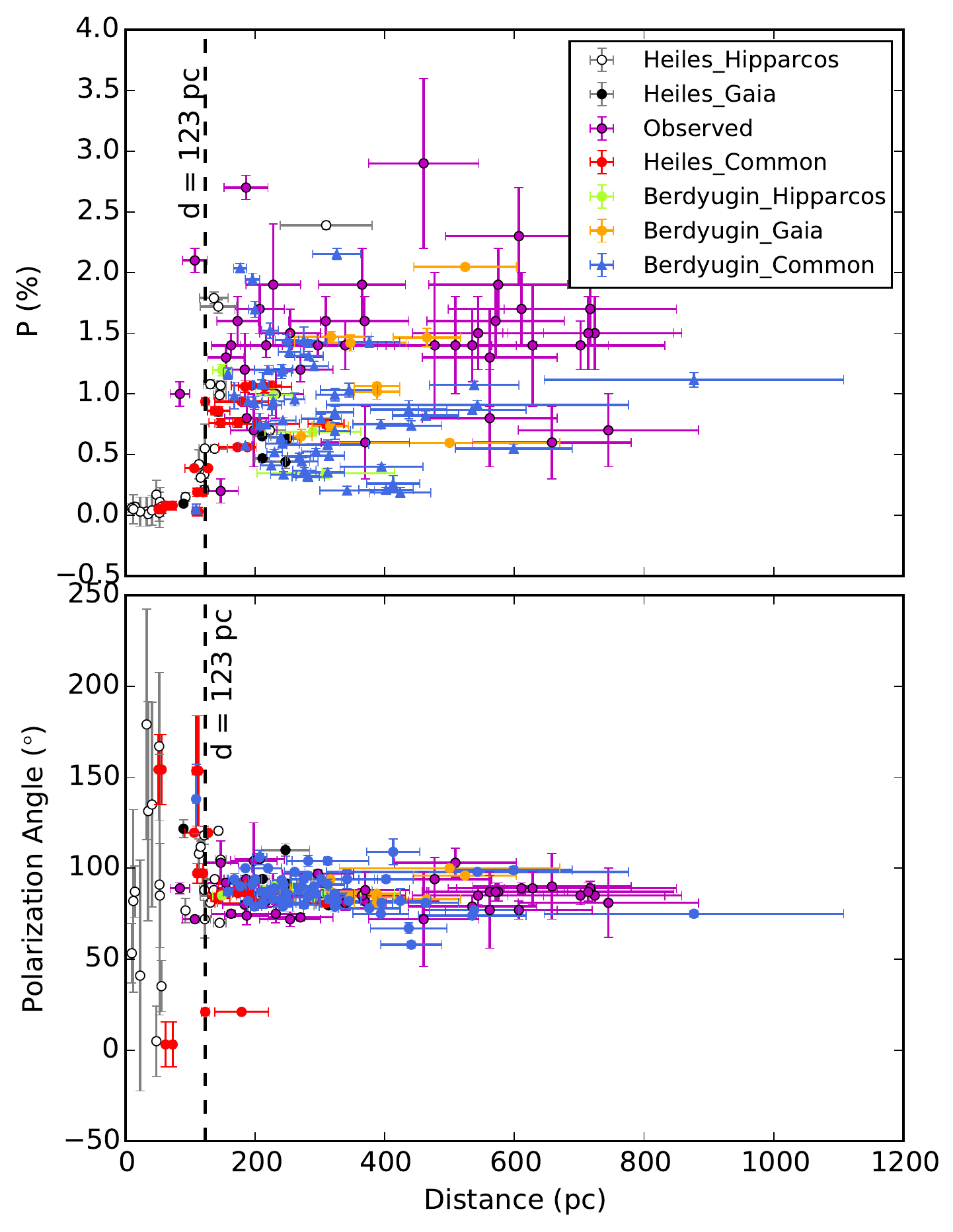}}
\caption{The degree of polarization (upper panel) and polarization position angle (lower panel) vs. distance of stars for which both polarization and distance measurements are available in the literature. The polarization and the parallax measurements are obtained from \citet{2000AJ....119..923H,2000A&A...358..717B} and \citet{2007A&A...474..653V}; \citet{2016yCat.1337....0G}, respectively. For the Heiles\_common and Berdyugin\_common stars, which have parallax measurements available in both the catalogs, we have taken the parallax values from Gaia.}\label{fig:heiles_hippP}
\end{figure}


We searched for stars with polarization measurements available in the catalogs by \citet{2000AJ....119..923H} and \citet{2000A&A...358..717B} within a region of $20\degree\times20\degree$ around MBM 33-39 complex. We obtained 51 stars for which polarization measurements are available in \citet{2000AJ....119..923H} and 76 stars from \citet{2000A&A...358..717B}. We obtained parallax measurements for these stars from \citet{2007A&A...474..653V} and \citet{2016yCat.1337....0G}. For 40 Heiles sources we obtained parallax measurements from \citet{2007A&A...474..653V} and for 19 Heiles stars, we obtained parallax measurements from \citet{2016yCat.1337....0G}. Of these, 12 stars were common in both. Similarly, 67 sources from \citet{2000A&A...358..717B} have parallax measurements in \citet{2007A&A...474..653V}, 72 stars have parallax measurements in \citet{2016yCat.1337....0G}. Of these, 63 stars were common in both. Only those stars are selected for which the ratio of the parallax and the error in the parallax is greater than or equal to 2. In Figure \ref{fig:heiles_hippP} we show the degree of polarization (above) and polarization position angle (below) vs. distance estimated from the \citet{2007A&A...474..653V} using open circles for Heiles stars and from the \citet{2016yCat.1337....0G} using filled circles in black for Heiles stars. The green-yellow circles represent the Berdyugin stars having distance from \citet{2007A&A...474..653V} and orange circles correspond to Berdyugin stars that have distance from \citet{2016yCat.1337....0G}. The Heiles stars that are common in both \citet{2007A&A...474..653V} and \citet{2016yCat.1337....0G} are identified using red, while the Berdyugin common stars are shown using blue color. The magenta circles represent the observed sources for which the distances have been determined using the procedure described in section \ref{sec:dist_Av}.

A significant increase in the values of the degree of polarization is found to occur at $\sim$120 pc (marked using dashed line). The dispersion in the values of the polarization position angle was also found to be relatively large for sources lying till $\sim$120 pc. Beyond this distance, the mean value of the polarization position angle is found to be 88$\degree$, which is very much similar to the mean value of the sources observed by us towards the MBM 33-39 cloud complex. Based on the 2MASS photometry \citep{2003yCat.2246....0C}, using the $J-H$ and $H-K_{s}$ colors (described briefly in section \ref{sec:dist_Av}), we estimated the distance and A$_{V}$ of stars for which we have polarization observations. We selected sources having photometric errors in $JHK_{s}\leq 0.035$. Of the 234 stars observed by us, we could estimate distance and A$_{V}$ values for 41 of them which follow all our selection criteria. Majority of the sources are lying beyond the $\sim$120 pc distance of the complex. The mean value of the polarization position angle of these 41 sources is found to be 85$\degree$ implying that the magnetic field of the MBM 33-39 cloud complex is similar to those present in the ambient interstellar medium.

\subsubsection{From the 2MASS photometry}\label{sec:dist_Av}
\begin{figure}
\centering
\resizebox{8cm}{8cm}{\includegraphics{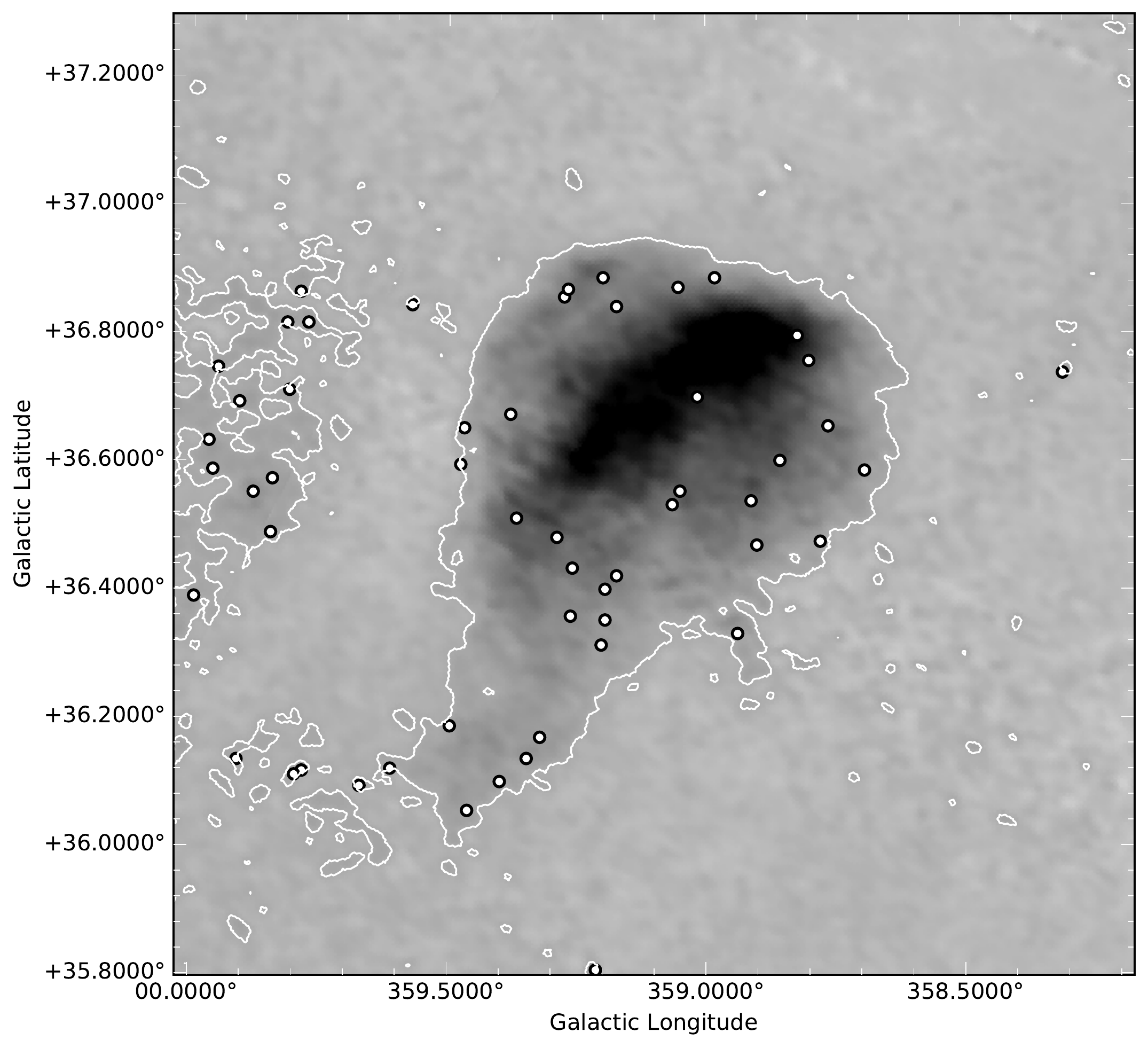}}
\caption{AKARI L-band (160 $\mu$m) image of MBM 33 cloud. The open circles show the stars lying inside the outermost contour of the image.}\label{fig:akari}
\end{figure}
\begin{figure}
\centering
\resizebox{8cm}{6cm}{\includegraphics{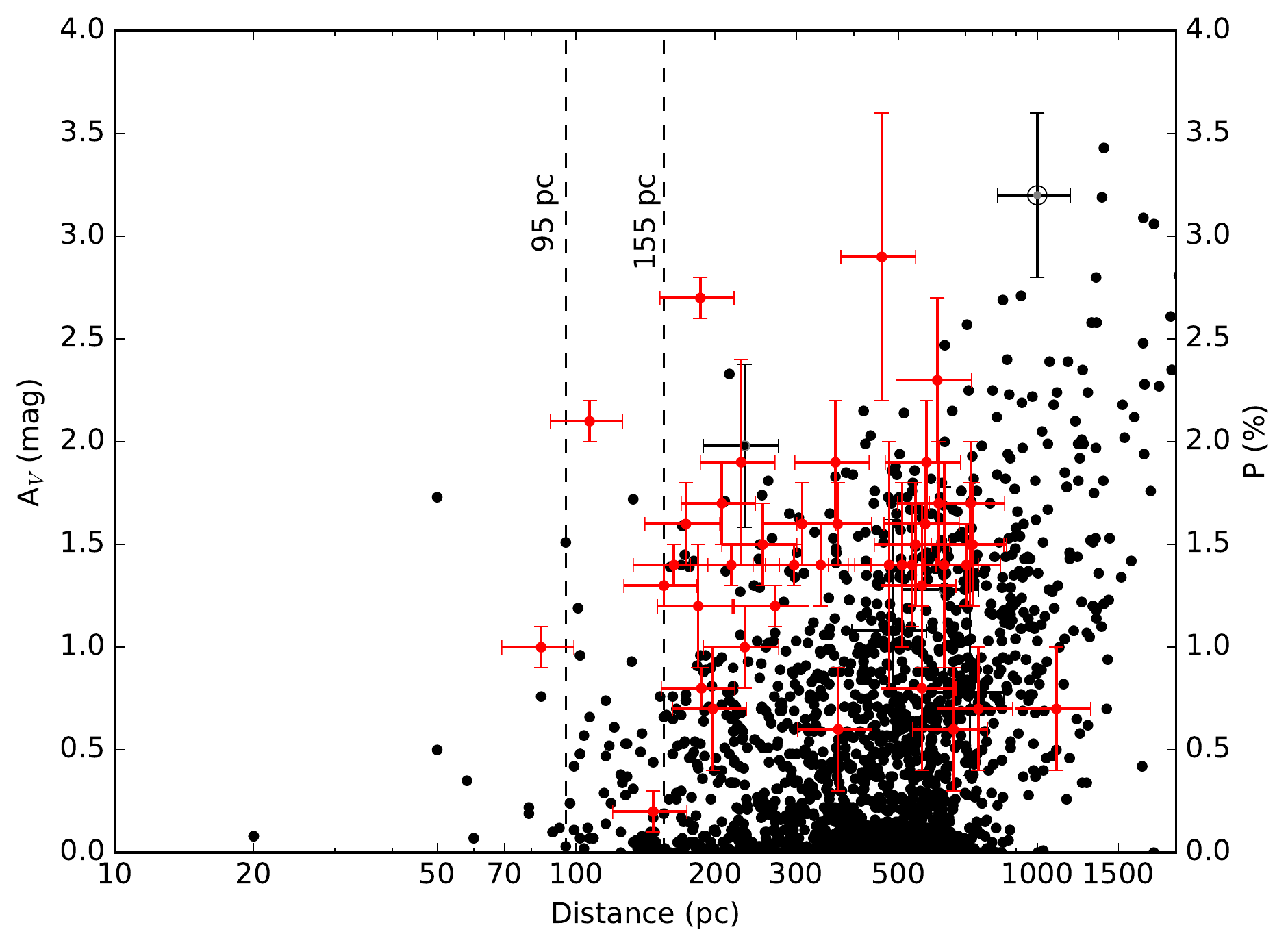}}
\caption{A$_{V}$ vs. distance plot for the stars (black circles) belong to the regions containing HLCs, MBM 33-39 complex. The dashed vertical lines are drawn at 95 pc and 155 pc inferred from the method described in \citet{2010A&A...509A..44M}. Typical error bars are shown on a few data points. The degree of polarization (P\%) is also plotted along second Y-axis using red circles.}\label{fig:dist_Av}
\end{figure}
Using the near-IR photometric method \citep{2010A&A...509A..44M}, we estimated the distance to the MBM 33-39 cloud complex. A short description of the method\footnote{For a more attentive analysis on the uncertainties and limitations of the method, refer \citet{2010A&A...509A..44M}} is described here. The method uses an approach in which the stars lying in the regions containing the clouds are classified into main sequence and giants using near-IR colors. The observed ($J-H$) and ($H-K_{s}$) colors of the stars having ($J-K_{s}$)$\leq0.75$ are de-reddened using trial values of $A_{V}$ simultaneously and plotted in ($J-H$) vs. ($H-K_{s}$) color-color (CC) diagram. The $JHK_{s}$ values are from the 2MASS catalog \citep{2003yCat.2246....0C}. We have considered only those sources for which the photometric errors in $JHK_{s}$ are $\leq 0.035$ and also are lying within the cloud boundary inferred from the 160 $\mu$m images of the clouds obtained by AKARI satellite. In Fig. \ref{fig:akari}, we identified sources selected based on the above criteria towards MBM 33 on the 160 $\mu$m image of the cloud. A normal interstellar extinction law \citep{1985ApJ...288..618R} is used for the de-reddening procedure. The spectral type corresponding to the minimum value of the $\chi^{2}$ produced by the best fit of the de-reddened colors to the intrinsic colors is assigned to the star. Thus, the stars, which are classified as main sequence stars, are plotted in an $A_{V}$ versus distance diagram to estimate the distance of the cloud. This method was used to estimate distances to a number of clouds \citep[e.g.,][]{2011A&A...536A..99M, 2013MNRAS.432.1502S, 2013A&A...556A..65E, 2016A&A...588A..45N}.

In Fig. \ref{fig:dist_Av}, we show the $A_{V}$ versus distance diagram for the sources selected towards all the seven MBM clouds using filled circles in black. The 41 sources for which we made polarization measurements and estimated their distance and $A_{V}$ using the 2MASS photometry are also plotted using filled circles in red. A significant increase ($\sim$1 magnitude) in the values of $A_{V}$ is apparent at $\sim95$ pc. A second increase in the $A_{V}$ ($\sim$2 magnitude) is occurring at $\sim155$ pc. The distance of the majority of the sources observed by us are also found to be lying beyond 155 pc and show higher values of $A_{V}$ ($\sim$2 magnitude).

Distance to interstellar material present in a region covering an area of 8$\degree\times$ 9$\degree$ towards MBM 33-39 complex was obtained by \citet{1989A&A...223..313F} using four color \textit{uvby} and H$\beta$ photometry of 81 stars which are of A and F spectral types. \citet{1989A&A...223..313F} assigned a distance of $110\pm10$ pc to all the interstellar material lying towards the region of study. \citet{2014ApJ...786...29S} presented a catalog of distances to molecular clouds in which they estimated distances to the majority of the clouds identified by \citet{1985ApJ...295..402M}. \citet{2014ApJ...786...29S} estimated distances to the molecular clouds using optical photometry of stars obtained from PanSTARRS-1. They used the technique of \citet{2014ApJ...783..114G} to obtain distance and reddening of individual stars. The distances obtained by \citet{2014ApJ...786...29S} are presented in Table \ref{HLC_para}. According to \citet{2014ApJ...786...29S}, the clouds in MBM 33-39 complex is lying in the range of $\sim$80$-$120 pc within the uncertainty of $\sim$10$-$30\%. 

The change in the values of the polarization position angles from being random to a more regular value of about 88$\degree$ is seen clearly in Fig. \ref{fig:heiles_hippP} for stars with distance $\gtrsim$120 pc. It is at this distance and beyond that, the values of the degree of polarization are also becoming significantly high. The results obtained from polarization measurements indicate that the high density material towards the MBM 33-39 complex is most likely lying at a distance of $\sim$120 pc with an uncertainty of 10\%.

\subsection{Magnetic field morphology of the region}

This work presents the first polarimetric results of a high galactic latitude cloud complex that contains both translucent as well as the dark molecular cloud \citep[][]{1995A&A...295..755T, 2009ApJ...701.1044K}. The initial study of interstellar polarization towards high galactic regions of north and south Galactic poles was made by, for example, \citet{1968ApJ...151..907A, 1970MmRAS..74..139M, 1979A&A....74..201M, 1986A&AS...64..487K, 2000A&A...358..717B, 2001A&A...368..635B, 2001A&A...372..276B, 2004A&A...424..873B}. \citet{2014A&A...561A..24B, 2011ASPC..449..157B} made polarization measurements of high galactic latitude ($30\degree<b<70\degree$) field stars that are lying in the longitude range of $240\degree<l< 360\degree$, $0\degree<l< 60\degree$ and $240\degree<l< 360\degree$. The main aim of their study was to make a comprehensive polarization map of the region and understand the patterns, and its relation to the local dust and gas shells and with the local spiral arm. The most remarkable feature seen in their study is a loop like structure centered at $l\approx330\degree$ which was noticed in previous studies also \citep{1978Ap&SS..54..425E, 1970MmRAS..74..139M}. It was suggested that the local super bubble or the Loop I \citep{1966MNRAS.131..335L, 1971A&A....14..252B} is responsible for the observed polarization pattern seen in their study. The MBM 33-39 complex considered in this study is lying in the longitude range of $359\degree<l< 360\degree$, $0\degree<l< 12\degree$ and latitude range of $35\degree<b<39\degree$. The magnetic field geometry of the MBM 33-39 complex traced in this study will be useful to understand the relationship between the large scale and individual cloud scale magnetic field orientation and their relationship with the material structures of the cloud. 
 
\begin{figure*}
\centering
\resizebox{\textwidth}{20.0cm}{\includegraphics{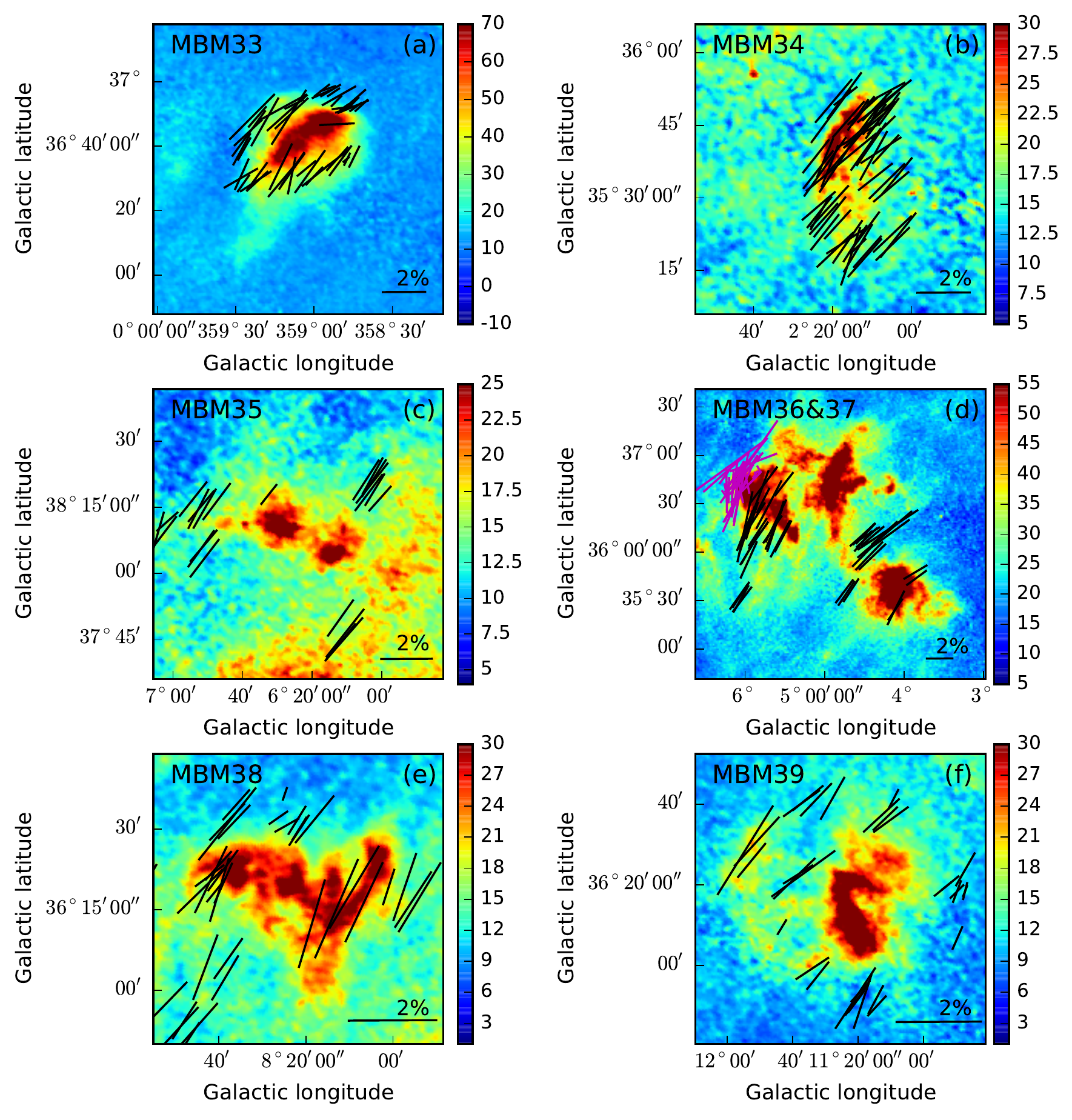}}
\caption{Polarization vectors (depicting the magnetic field morphology) in black color drawn on the AKARI 160 $\mu$m images of seven MBM clouds are shown in (a), (b), (c), (d), (e), (f), respectively. The NIR polarization vectors for MBM 37, taken from \citet{2012ApJ...748...18C}, are also plotted in magenta color.}\label{fig:mbmcomplex}
\end{figure*}
Fig. \ref{fig:mbmcomplex} shows the polarization vectors of MBM 33, 34, 35, 36, 37, 38 and 39 superposed on their corresponding AKARI Infrared Astronomy Satellite 160 $\mu$m images. The 160 $\mu$m image shows the intensity of the cold dust emission from the clouds. In Fig. \ref{fig:mbmcomplex} (a) we present the polarization vectors of 49 stars observed towards MBM 33. The cloud shows a cometary shape with the head oriented away from the Galactic plane. The high density part of the cloud is oriented along the north-west and south-east direction of the main body of the cloud. The magnetic field lines inferred from the polarization vectors are also oriented parallel to this high density structure of the cloud. Among the seven clouds studied here, MBM 33 shows the largest value of dispersion in polarization position angles (15$\degree$).

In Fig. \ref{fig:mbmcomplex} (b) we show the polarization vectors for MBM 34, which lies about 3$\degree$ east of MBM 33. The 160 $\mu$m emission shows a curved cloud structure as we move from the north towards east. Towards the north-eastern edge, the polarization vectors seem to follow the curvature of the cloud though, in general, the magnetic field is oriented parallel to the global field direction. The polarization vectors towards MBM 35 is shown in the  Fig. \ref{fig:mbmcomplex} (c). The cloud shows two relatively dense condensations. A line joining these two structures is oriented almost perpendicular to the mean direction of the magnetic field lines. The two condensations could have been formed as a result of fragmentation of the cloud essentially occurred perpendicular to the direction of the mean magnetic field. The emission from the low density material in the south-western parts of the two condensations is distributed almost parallel to the magnetic field orientation.

In Fig. \ref{fig:mbmcomplex} (d) we show the polarization vectors for MBM 36 and MBM 37. MBM 36 is also identified as LDN 134 \citep{1962ApJS....7....1L}. It shows a spherical shape. Projected magnetic field inferred from the optical polarization measurements shows a smooth distribution. The magnetic field is well aligned with the global magnetic field mapped by \citet{2011ASPC..449..157B, 2014A&A...561A..24B}. MBM 37 (= LDN 134N, LDN 183) harbors a starless centrally condensed ($\sim$180 visual magnitude), high column density prestellar core. This core has been studied extensively in different spectroscopic species \citep{2000ApJ...542..870D, 2001ApJS..136..703L, 2005A&A...429..181P, 2007A&A...467..179P} and in far-infrared and sub-millimeter continuum \citep{2002A&A...382..583J, 2002MNRAS.329..257W, 2003A&A...398..571L, 2003A&A...406L..59P, 2005MNRAS.360.1506K, 2008A&A...487..993K}. This core is considered to be much more evolved than many other prestellar cores \citep{2000ApJ...542..870D} and also shows ample evidence of it being in the verge of initiating star formation \citep{2001ApJS..136..703L}. Three spatially and kinematically distinct sub-cores have been identified towards the densest part of LDN 183 \citep{2009ApJ...701.1044K}. Based on the results obtained, \citet{2009ApJ...701.1044K} proposed that LDN 183 is a prestellar core which is currently fragmenting, collapsing and rotating about its inner magnetic field inferred from sub-mm observations. The core is spinning up as it collapses and may form a multiple protostellar systems.

\begin{figure}
\centering
\resizebox{8.7cm}{10cm}{\includegraphics{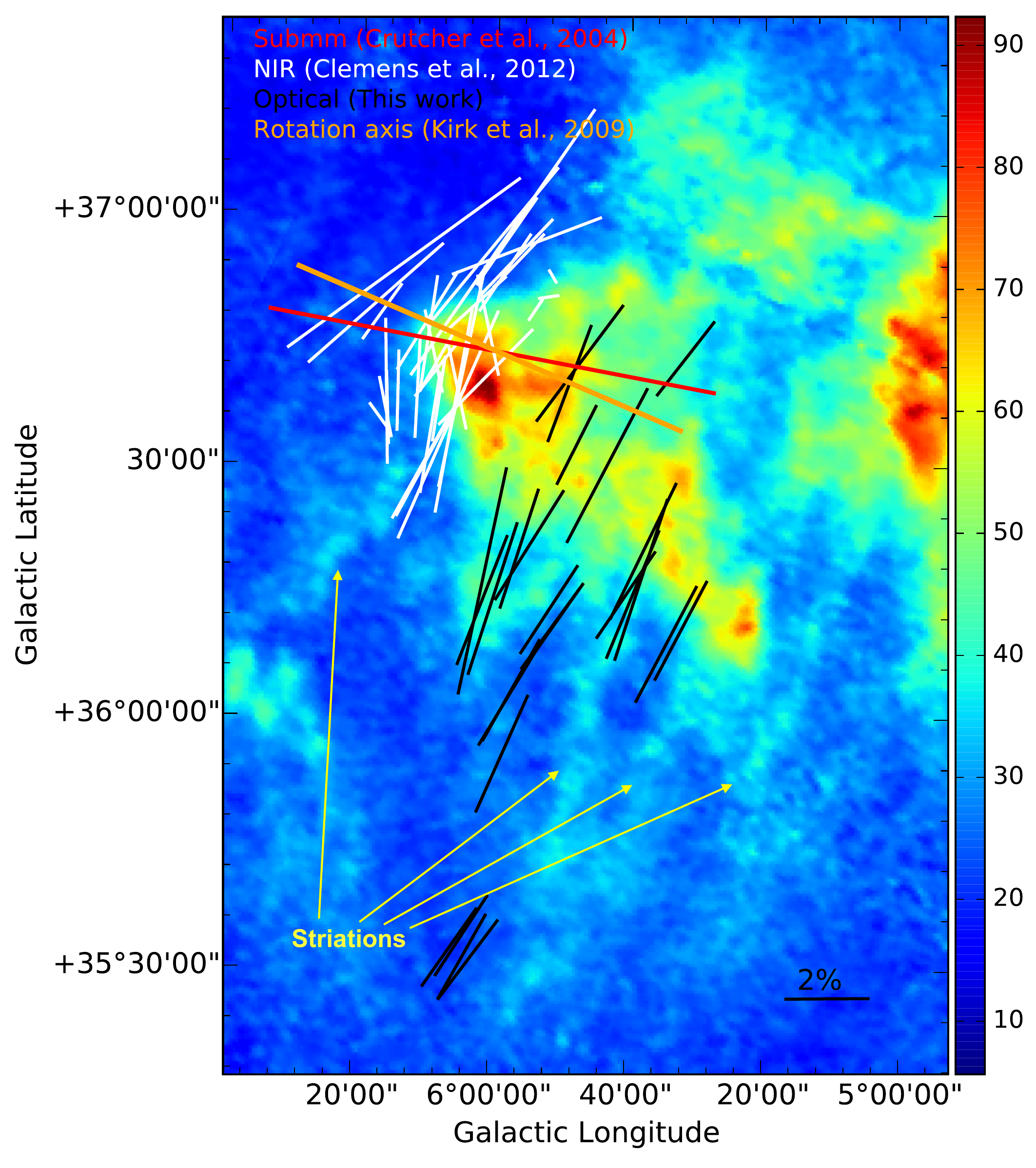}}
\caption{AKARI 160 $\mu$m images of MBM 37. The observed optical polarization vectors are plotted in black color, NIR polarization vectors, taken from \citet{2012ApJ...748...18C}, are plotted in white color. The red vector shows the magnetic field orientation inferred from the averaged sub-mm polarization vector corrected for 90$\degree$ \citep{2004Ap&SS.292..225C}. The orange vector represents the cloud's rotation axis \citet{2009ApJ...701.1044K}.} \label{fig:l183}
\end{figure}

\begin{figure*}
\centering
\resizebox{14cm}{13cm}{\includegraphics{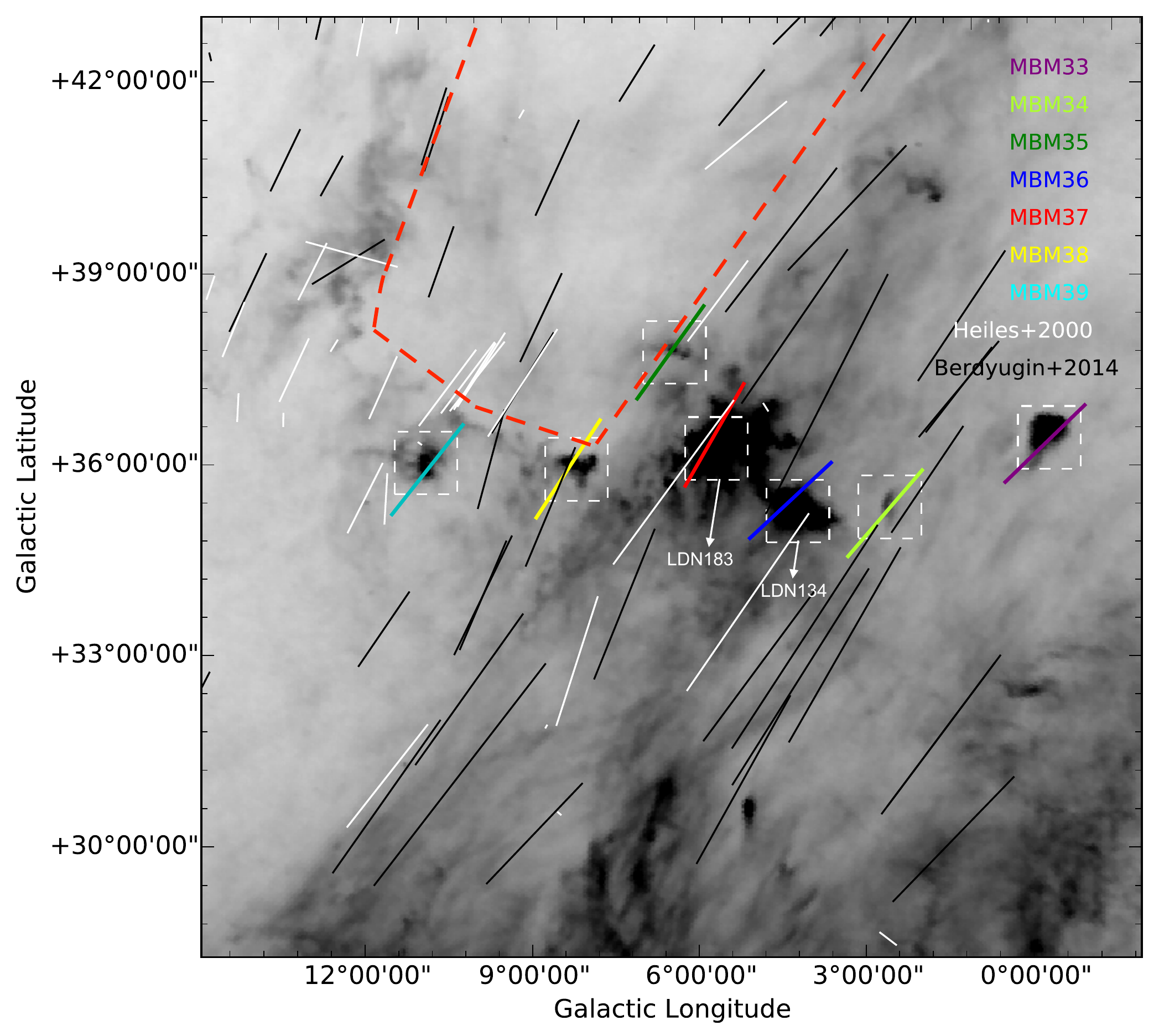}}
\caption{15$\degree\times$15$\degree$ Planck 857 GHz image of MBM 33-39 cloud complex. The observed mean polarization vectors corresponding to each cloud are overplotted. The white and black polarization vectors, taken from \citet{2000AJ....119..923H} and \citet{2000A&A...358..717B}, respectively, are also overplotted. White boxes identify each cloud. A peculiar U-shape structure is shown by dashed-line in red.}\label{fig:mbm_tot}
\end{figure*}

In Fig. \ref{fig:l183} we show a close-up view of LDN 183 region. The black vectors represent the polarization measurements made by us. The near-infrared polarization vectors measured by \citet{2012ApJ...748...18C} are plotted in white. \cite{2004ApJ...600..279C} made SCUBA\footnote{SCUBA camera is mounted on JCMT, which is operated by the Joint Astronomy Center, Hawaii, on behalf of the UK PPARC, the Netherlands NWO, and the Canadian NRC.} polarization measurements around LDN 183 and observed 25 independent positions. The mean value of all the sub-mm polarization position angles corrected for the 90$\degree$ rotation to infer the projected magnetic field orientation in the inner high-density region of LDN 183 is shown using a red line. The rotation axis identified by \citet{2009ApJ...701.1044K} is shown using an orange line. The outer projected magnetic field inferred from the optical polarization measurements is well aligned with the field orientation of the loop structure. The long axis of the cloud is found to be roughly perpendicular to the outer magnetic field projected on the sky plane suggesting that the accumulation of the material occurred along the magnetic field lines. Based on $^{12}$CO emission, \citet{2008ApJ...680..428G} showed the striation patterns that are aligned with the local magnetic field which suggests a strong coupling between the magnetic field and the ISM. In LDN 183 also we see such patterns in the low density regions (Fig. \ref{fig:l183}) that are found to be aligned with the local magnetic field inferred from our polarization vectors. These striations are either due to the Kelvin-Helmholtz instability or magnetosonic waves propagating through the envelope of the cloud \citep{2016MNRAS.461.3918H}. The high density part of the cloud where near-infrared polarization measurements are made by \citet{2012ApJ...748...18C}, the vectors follow the curvature of the cloud. It is interesting to note that the curved part of LDN 183, MBM 38 and MBM 39 together they form part of a peculiar structure identified using a dotted line in Fig. \ref{fig:mbm_tot}. The rotation axis of LDN 183 \citep{2009ApJ...701.1044K} is oriented parallel to the inner magnetic field but lying perpendicular to the outer magnetic field (all the quantities are in the sky plane). The projected field lines inferred from sub-mm and optical/near-infrared are found to be oriented perpendicular to each other. This has also been noticed in our previous study towards IRAM04191+1522 \citep{2015A&A...573A..34S}. Further, the polarization vectors obtained from optical observations are almost perpendicular to the long axis of LDN 183 while the polarization vector obtained from submillimeter observations is roughly parallel to the long axis of L183. Since, the former likely traces the densest part of LDN 183 while the later likely traces the diffuse part of LDN 183, this may mean that the inner region of LDN 183 is close to the supercritical status while the outer region of L183 is in subcritical status.

The polarization vectors towards MBM 38 and MBM 39 are shown in the  Fig. \ref{fig:mbmcomplex} (e, f). MBM 38 lies on the periphery of the structure identified with a broken line in red on Fig. \ref{fig:mbm_tot}. The plane of the sky magnetic field lines in both MBM 38 and 39 are found to be distributed smoothly. In MBM 38, the long axis of the cloud is found to be roughly perpendicular to the magnetic field lines. In MBM 39, similar to MBM 33 and 34, the material structure shows a curved orientation which is not apparent in the magnetic field map owing to a few number of stars observed which are projected over the structure. 

\cite{2016A&A...586A.135P} studied the MBM 33-39 cloud complex and mapped sub-mm magnetic field morphology around this region \citep[see Fig. 18 in ][]{2016A&A...586A.135P}. They found that the magnetic field geometry is aligned with the matter at high Galactic latitudes in diffuse ISM. Our optical magnetic field map is found to be consistent with the Planck sub-mm magnetic field map, where polarization is caused by the emission from dust grains. The large-scale magnetic field structure shown in Fig. 8 is also found to be consistent with all-sky Planck sub-mm magnetic field map \citep[see Fig. 5 in ][]{2015A&A...576A.104P}.

\subsection{Magnetic field strength}

We estimated the plane-of-the-sky component of magnetic field strength (B$_{pos}$) in HLCs using the updated Chandrashekhar-Fermi (CF) relation \citep{1953ApJ...118..113C, 2001ApJ...546..980O, 2005mpge.conf..103C}:
\begin{equation}\label{eq3}
B_{pos} = 9.3\sqrt{\mathrm{n(H_{2})}} \frac{\Delta \mathrm{V}}{\delta\theta}  
\end{equation} 

In this equation, n(H$_{2}$) represents the number density of the molecular hydrogen in molecules cm$^{-3}$, $\Delta$V is the full-width half maximum of CO line in km s$^{-1}$ and $\delta\theta$ shows the dispersion in polarization position angle $\theta_{P}$ in degrees. \citet{1953ApJ...118..113C} assumed an equipartition between the kinetic and perturbed magnetic energies and it was suggested that the plane-of-sky component of magnetic field strength could be estimated using together with the velocity dispersion and polarization position angle dispersion. For molecular clouds, MBM 33, MBM 36, MBM 37, MBM 38 and MBM 39, we obtained the number density and the CO line width ($\Delta$V) values from \citet{1995ApJS..101...87L} and for MBM 34, we used the mean value of all the number densities and the $\Delta$V value is taken from \citet{1998ApJ...492..205H}. The values of $\delta\theta$ for all the clouds are estimated from the standard deviation in polarization angle $\theta_{P}$. The $\delta\theta$ values are corrected for the uncertainties in polarization angle $\theta_{P}$ \citep{2001ApJ...561..864L, 2010ApJ...723..146F}. 
Using these values in equation \ref{eq3}, we estimated the value of B$_{pos}$. The values of number density, $\delta\theta$, $\Delta$V and magnetic field strength for the six clouds are listed in Table \ref{tab:magnetic_field}. The strength of the magnetic field range from $\sim$10$-$40 $\mu$G. The strongest field strength is found in LDN 183 (MBM 37) and the weakest field is found in MBM 33. 

\begin{table*}
\centering
\caption{The magnetic field strength in HLCs.}\label{tab:magnetic_field}
\begin{tabular}{cccccc}\hline
Cloud    & Number Density & $\delta\theta$ & $\Delta V$ & B  & Reference\\
Name  	 & (cm$^{-3}$)    & ($\degree$)   & (km s$^{-1}$) & ($\mu$G) & \\\hline
MBM33    &  990  & 15  & 0.55   & 11 & a \\
MBM34    &  966	 & 9   & 0.70   & 22 & b \\
MBM36    &  905  & 9   & 1.04   & 32 & a \\
MBM37	 &  878  & 8   & 1.23   & 42 & a \\
MBM38	 &  835  & 12  & 0.54   & 12 & a \\
MBM39	 & 1220  & 13  & 0.58   & 14 & a \\\hline
\end{tabular}\\
a - \cite{1995ApJS..101...87L} \\
b - \cite{1998ApJ...492..205H}
\end{table*}

\subsection{Possible influence of local environment and magnetic field in the formation of MBM 33-39 complex}
\begin{figure*}
\centering
\resizebox{17cm}{11cm}{\includegraphics{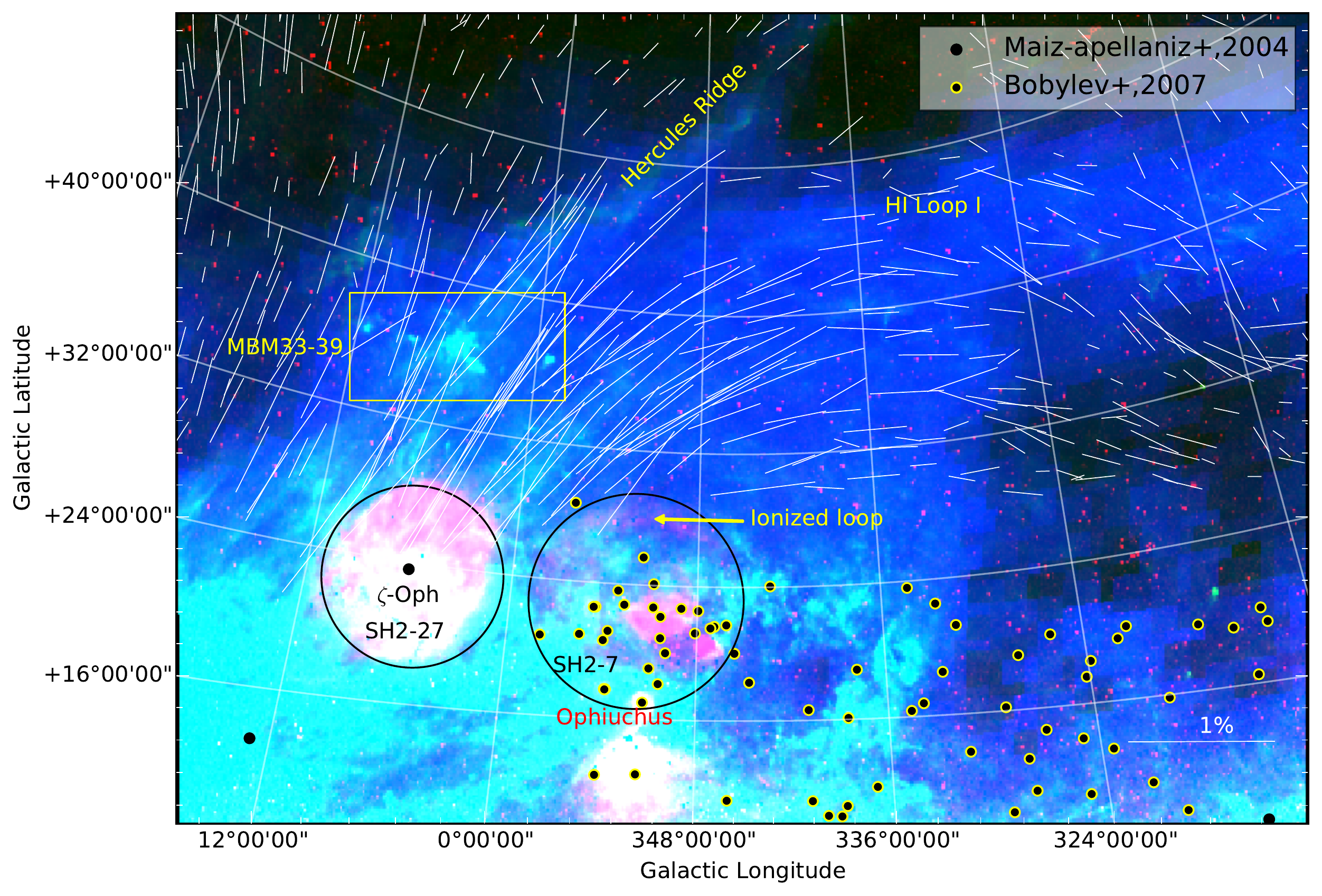}}
\caption{The 70$\degree\times$70$\degree$ color composite image of the Sco-Cen region which contain MBM 33-39 cloud complex centered at $l=345\degree$ and $b=+25\degree$. The red, green and blue colors correspond to H$\alpha$ image, 857 GHz Planck image and nH image obtained from the SkyView. The polarization vectors in white, taken from \citet{2000A&A...358..717B}, are also overplotted. A vector with 1\% polarization is shown as scale.}\label{fig:color_composite}
\end{figure*}

In order to examine the kind of environment prevails in the close vicinity of MBM 33-39 complex, we made a color composite image of the region containing the MBM 33-39 cloud complex. In Fig. \ref{fig:color_composite} we show the color composite image of 70$\degree\times70\degree$ region centered at $l=335\degree$ and $b=+20\degree$. It is created using H$\alpha$ image (red), 857 GHz Planck image (green) and atomic hydrogen column density map (blue) all obtained from the SkyView\footnote{SkyView has been developed with generous support from the NASA AISR and ADP programs (P.I. Thomas A. McGlynn) under the auspices of the High Energy Astrophysics Science Archive Research Center (HEASARC) at the NASA/ GSFC Astrophysics Science Division.}. The red color shows the distribution of the ionized gas in this region. The ionized region is primarily distributed around the $\zeta$ Ophiuchus \citep[Sh2-27, ][]{1959ApJS....4..257S} and the Sco OB2 \citep[Sh2-7, ][]{1959ApJS....4..257S} association which are identified and labeled. Some of the members of the Scorpius-Centaurus association (LCC = Lower Centaurus-Crux, UCL = Upper Centaurus-Lupus, and US = Upper Scorpius) obtained from \citet{2007AstL...33..571B} are identified using black-yellow circles. Distribution of O-type stars towards displayed region obtained from \citet{2004ApJS..151..103M} is also shown using filled black circles. The location of the MBM 33-39 cloud complex is identified using a rectangular box. The polarization vectors measured by \citet{2014A&A...561A..24B} are also over-plotted for better visualization. 

Clearly, there are a number of loop structures visible in the Fig. \ref{fig:color_composite}. A small-scale loop structure of ionized gas is noticeable towards the northern part of the Sh2-7 which is followed by another bigger loop of HI gas, Loop I \citep{1966MNRAS.131..335L}. There is one more loop which is identified with Hercules ridge \citep{1973A&A....24....1F}. Both the bigger loops are conspicuous in the 857 GHz Planck image which traces the distribution of cold interstellar dust grains and in the polarization map produced by \citet{2014A&A...561A..24B}. According to the basic concept of star formation history of the Scorpius-Centaurus association \citep{1992A&A...262..258D, 1995A&A...295..755T, 1999AJ....117.2381P}, the star formation process started in the UCL association some 15 Myr ago. The most massive member of UCL exploded as a supernova about 12 Myr ago, which formed most of the HI shells in the vicinity of the association. As the shock wave produced by this supernova passed through US some 5 Myr, it triggered star formation there. Stellar winds from the massive members of US initiated to spread the molecular cloud and paused the star formation process. About 1.5 Myr ago, the most massive member in the US went off as a supernova which fully dispersed the parent molecular cloud. The passage of the shock wave about 1 Myr ago has triggered recent star formation in $\rho$ Oph. Thus there has been at least two events of supernova explosions in the vicinity of the location where MBM 33-39 cloud complex is situated. A link between the Loop I and the runaway star $\zeta$ Ophiuchus was suggested by \citet{1971A&A....14..252B}. The distance of $\zeta$ Ophiucus corresponding to a parallax of 8.91 (0.2) mas \citep{2007A&A...474..653V} is $\sim110\pm3$ pc. Thus $\zeta$ Ophiuchus is about 35 pc from the MBM 33-39 complex and hence MBM 33-39 cloud complex might be currently interacting with the stellar wind and the ionizing radiation from $\zeta$ Ophiuchus also. These interactions probably might have driven the large-scale flows of material, guided by the large-scale magnetic field, that is required for the initiation of cloud formation.
The alignment between the matter structure in the ISM and the magnetic field shows evidence of the formation of cold neutral filaments through turbulence. The condensation of cold gas out of warm neutral medium could be triggered by the local environment \citep{2005A&A...433....1A, 2009ApJ...704..161I, 2009ApJ...695..248H, 2014A&A...567A..16S}. Scenario in which if the gas velocity is dynamically aligned with the magnetic field, the gas condensations tend to get stretched due to turbulent shear creating filaments and sheets, which would emerge elongated in column density maps. On the other hand, if the velocity shear stretches the matter into filaments, due to flux freezing, the field also gets stretched creating alignments between the magnetic field and the cold condensations  \citep{2013ASPC..476..115H}. In MBM 33 and 34, the structures seem to be aligned with the projected magnetic field. In MBM 35, though the high density structure is aligned perpendicular to the projected magnetic field, the diffused emission seems to be parallel to the field. 

Based on a quantity which is a normalized velocity difference of peak velocities of optically thick and thin lines, \citet{2011ApJ...734...60L} have classified MBM 37 (LDN 183) as a contracting core. When a gravitationally bound structure forms, an equipartition between gravitational and turbulent energies are expected and turbulence becomes super-Alfv\'{e}nic (sub-Alfv\'{e}nic) for supercritical (sub-critical) structures. For sub-critical structures, the magnetic field is dynamically important and draws matter preferentially parallel to the field lines. This results in the formation of sheet-like structures which subsequently get fragmented to form elongated filaments. The presence of striations perpendicular to the high-column density ridge of MBM 37 supports this picture of cloud formation. Such striations perpendicular to the high-column density filaments have been noticed in the Herschel maps of nearby molecular clouds, like Taurus \citep{2013A&A...550A..38P}. The magnetic field strength estimated for the MBM 33-39 clouds are typical of low to moderate strength. Though the projected envelop magnetic field inferred from the optical polarization is consistent with the standard model of cloud collapse, almost orthogonal orientation of the projected magnetic field at the inner regions of MBM 37 inferred through sub-mm polarization \citep{2004ApJ...600..279C} is puzzling. The rotation axis of the cores identified by \citet{2009ApJ...701.1044K} is also found to be aligned with this inner magnetic field. Magnetic field orientation of more fainter stars need to be made to obtain a much clear picture of the role played by the magnetic field in the formation of first translucent and then dark high density clouds in ISM.

\subsection{Serkowski's law} 
\begin{figure*}
\centering
\resizebox{15.5cm}{14.5cm}{\includegraphics{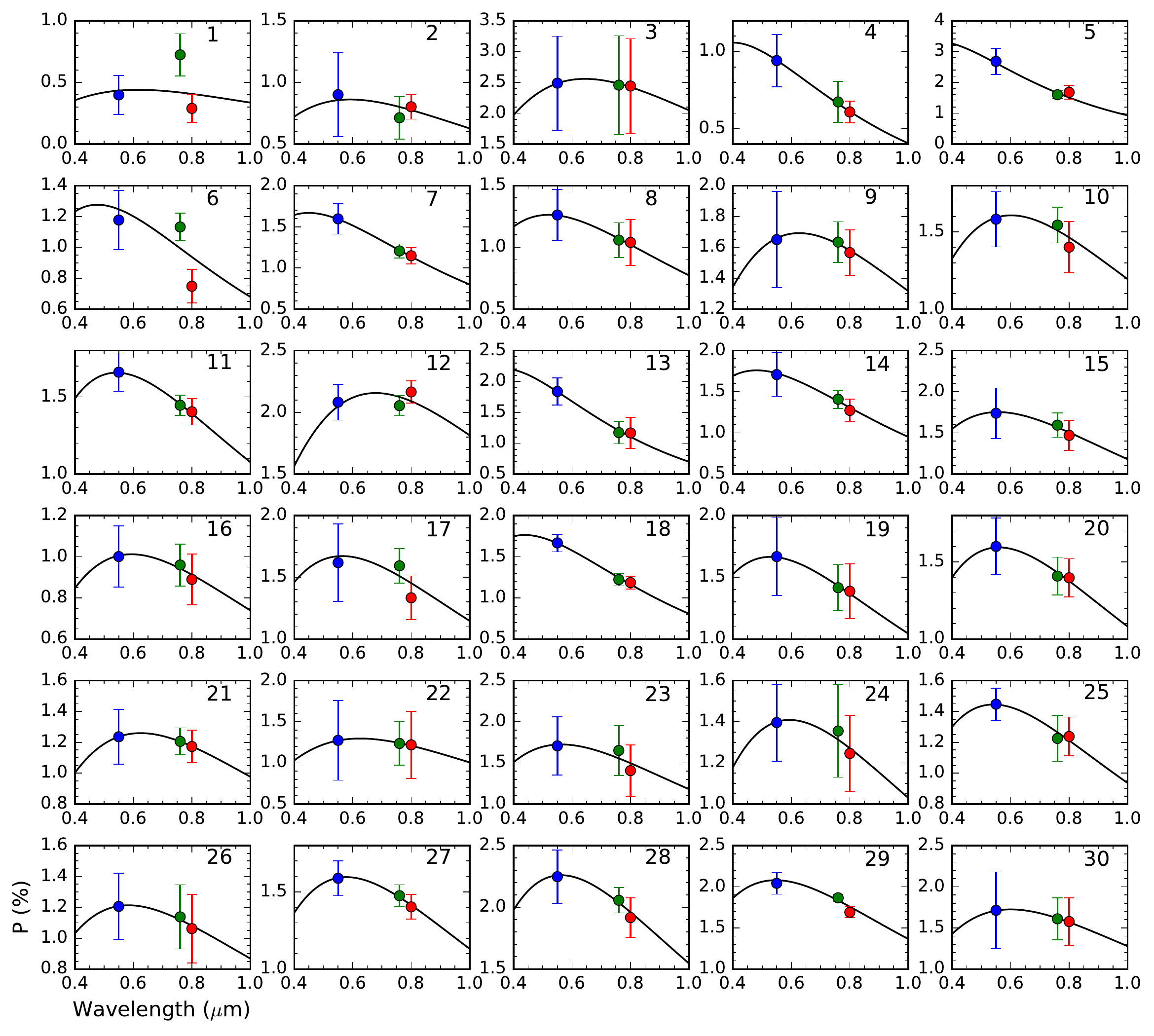}}
\caption{\small Interstellar linear polarization curves for 30 stars with the wavelength fitted using Serkowski's law \citep{1975ApJ...196..261S}. Blue, green and red circles are observed polarization values for V, R and I filters, respectively. Fitting to each group of VRI points has been shown. The star IDs corresponding to table \ref{tab:VRI} has been shown in each panel.}\label{fig:Serkowski1}
\end{figure*}

The value of P strongly depends on the wavelength, which is known as the polarization curve, described by the Serkowski law \citep{1975ApJ...196..261S} as follows:

\begin{equation}\label{eq:Serkowski}
P_\lambda = P_{max}~exp\Big[-K~ln^2\Big(\frac{\lambda_{max}}{\lambda}\Big)\Big].
\end{equation}
Here, $P_{\lambda}$ is the degree of polarization in percentage at wavelength $\lambda$ and $P_{\mathrm{max}}$ is the maximum degree of polarization in percentage at wavelength $\lambda_{\mathrm{max}}$. The parameter $K$ which was originally taken to be fixed at a value of $K=1.15$, which determines the width of the peak in the curve. We studied the dust properties of MBM clouds using the Serkowski's relation (equation \ref{eq:Serkowski}). The equation \ref{eq:Serkowski} provides the relation between the maximum degree of polarization P$_{max}$ and corresponding wavelength $\lambda_{max}$ for each star. We calculated the P$_{max}$ and $\lambda_{max}$ using the weighted least-squares fitting to the measured polarization in V, R and I bands to equation \ref{eq:Serkowski} by assuming K = 1.15. 
The origin of the polarization can also be inferred by the $\lambda_{max}$ values. The stars having much lower $\lambda_{max}$ values than the mean value of the interstellar medium \citep[0.55 $\mu$m;][]{1975ApJ...196..261S} may possess an intrinsic component of polarization. Fig. \ref{fig:Serkowski1} shows the Serkwoski fitting for 27 stars towards MBM 33, MBM 34 and MBM 35. We also plotted their Serkowski fitting curve in the same figure. In the figure, two stars having star IDs 1 and 6, have one of the data points lying outside the best-fitted curve within errors. Moreover, for stars with star IDs 1 and 22, the ratio of the degree of polarization and polarization uncertainty (P$_{V}$/$\epsilon_{P_{V}}$) is less than 3. Therefore, we have excluded these three stars. The weighted average values of the P$_{max}$ and $\lambda_{max}$ with weighted standard deviations for remaining 27 stars are found to be 1.5 $\pm$ 0.4 $\%$ and 0.54 $\pm$ 0.08 $\mu$m, respectively. 


The estimated $\lambda_{max}$ is quite similar to the value corresponding to the general interstellar medium \citep[0.55 $\mu$m;][]{1975ApJ...196..261S} within error limits. We estimated the value of R$_{V}$, the total-to-selective extinction, using the relation R$_{V}$ = (5.6 $\pm$ 0.3) $\times$ $\lambda_{max}$ \citep{1978A&A....66...57W}, for all the stars. The R$_{V}$ values range from 1.95 to 3.80 for all the stars. The weighted average value of R$_{V}$ for 27 stars is found to be 2.9 $\pm$ 0.5, which is consistent with the mean value (R$_{V}$ = 3.08) for the Milky Way Galaxy, suggesting that the dust grain size within the MBM 33, MBM 34 and MBM 35 clouds is normal. The error in R$_{V}$ has been estimated by error propagation in the above equation. Table \ref{tab:VRI} shows the fitting parameters P$_{max}$, $\lambda_{max}$ and R$_{V}$ values for all the stars.

\section{Conclusion}\label{sec:conclude}

We made polarimetric observations of 234 stars that are located towards the direction of a high galactic latitude cloud complex, MBM 33-39. The complex contains clouds that are of both translucent and dark molecular cloud types. Thus, the study of this complex provides us a unique opportunity to understand the role played by the magnetic field in the early stages of cloud formation mechanism. A summary of the important results obtained from this study is listed below:

1. By combining our polarization results and those from the literature, we estimated distance to the complex. The distances of the individual stars are determined using the parallax measurements made by Hipparcos and/or GAIA. We estimated a distance of $\sim120\pm10$ pc for the complex. The advantage of using polarization results and parallax measurements is that the distance determination is independent of the properties of the observed stars.

2. The magnetic field geometry of the individual clouds inferred from our polarimetric results reveals that the field lines are in general consistent with the global magnetic field geometry of the region obtained from the previous studies. This implies that the clouds in the complex are permeated by the interstellar magnetic field.

3. The dust structures of the individual clouds inferred from 160 $\mu$m images from AKARI are found to be either parallel or perpendicular to the magnetic field suggesting that magnetic field played an important role in the formation process of these structures in the cloud.

4. The dispersion of polarization angles is found to be maximum in the MBM 33 suggesting that the cloud is in a state of dynamical evolution possibly due to its interaction with external triggers.

5. In MBM 37 (=LDN 183), the most evolved cloud in the complex, the magnetic field direction inferred from our optical polarization measurement which traces field lines of the periphery of the cloud is found to be oriented almost orthogonal to the field directions of the denser inner regions of the cloud traced by the sub-mm polarization measurements.

6. Multi-wavelength polarization measurements of a few stars projected on the clouds suggest that the dust grain size in these clouds is similar to those found in the normal interstellar medium of the Milky Way.

7. A possible formation scenario of MBM 33-39 complex is discussed by combining the polarization results from our study (and from the literature) and by identifying the distribution of ionized, atomic and molecular (dust) components of material in the region.

\section*{Acknowledgements}
The authors thank anonymous referee for the helpful and constructive comments. This research has made use of the SIMBAD database, operated at CDS, Strasbourg, France. We also acknowledge the use of NASA's \textit{SkyView} facility (http://skyview.gsfc.nasa.gov) located at NASA Goddard Space Flight Center. This publication makes use of data products from the Two Micron All Sky Survey (2MASS), which is a joint project of the University of Massachusetts and the Infrared Processing and Analysis Center/California Institute of Technology, funded by the National Aeronautics and Space Administration and the National Science Foundation.
CWL was supported by Basic Science Research Program though the National Research Foundation of Korea (NRF) funded by the Ministry of Education, Science. NS acknowledge Dr. A. K. Pandey (ARIES, Nainital, India) for his support. NS thanks Suvendu Rakshit (SNU, Seoul, Republic of Korea) for carefully reading the manuscript.

\bibliographystyle{mnras}
\bibliography{MBM34ref}
\newpage


\begin{sidewaystable}
    \vskip 18cm
\centering
\caption{Polarization values of observed Tycho stars in VRI filters.}\label{tab:VRI}
\begin{tabular}{cccccccccccc}\hline
Star ID & $\alpha$ (J2000) &  $\delta$ (J2000) & P$_{V}$ $\pm$ $\epsilon_{P_{V}}$ & $\theta_{V}$ $\pm$ $\epsilon_{\theta_{V}}$  &P$_{R}$ $\pm$ $\epsilon_{P_{R}}$ & $\theta_{R}$ $\pm$ $\epsilon_{\theta_{R}}$  & P$_{I}$ $\pm$ $\epsilon_{P_{I}}$ & $\theta_{I}$ $\pm$ $\epsilon_{\theta_{I}}$ & $\lambda_{max}$ $\pm$ $\epsilon_{\lambda_{max}}$ & P$_{max}$ $\pm$ $\epsilon_{P_{max}}$ & R$_{V}$ $\pm$ $\epsilon_{R_{V}}$ \\ 
  &($\degree$)&($\degree$)& (\%) 			&($\degree$)  & (\%) 			&($\degree$)  & (\%) 			&($\degree$) & (\%) & ($\mu$m)  & \\\hline
\multicolumn{12}{c}{\bf MBM 33} \\
1	 & 	234.666229	 & 	-7.290575	 & 	0.4 $\pm$ 0.2	 & 	51 $\pm$ 9	 & 	0.7 $\pm$ 0.2	 & 	78 $\pm$ 6	 & 	0.3 $\pm$ 0.1	 & 	71 $\pm$ 8	 & 	0.62 $\pm$ 0.67	 & 	0.44 $\pm$ 0.24	 & 	3.45 $\pm$ 3.75 \\ 
2	 & 	234.689987	 & 	-7.249656	 & 	0.9 $\pm$ 0.3	 & 	82 $\pm$ 4	 & 	0.7 $\pm$ 0.2	 & 	80 $\pm$ 6	 & 	0.8 $\pm$ 0.1	 & 	81 $\pm$ 3	 & 	0.59 $\pm$ 0.17	 & 	0.86 $\pm$ 0.16	 & 	3.31 $\pm$ 0.96 \\ 
3	 & 	234.697296	 & 	-7.267688	 & 	2.5 $\pm$ 0.8	 & 	60 $\pm$ 8	 & 	2.5 $\pm$ 0.8	 & 	65 $\pm$ 8	 & 	2.4 $\pm$ 0.8	 & 	70 $\pm$ 9	 & 	0.64 $\pm$ 0.01	 & 	2.56 $\pm$ 0.02	 & 	3.61 $\pm$ 0.20 \\ 
4	 & 	234.747787	 & 	-7.236612	 & 	0.9 $\pm$ 0.2	 & 	62 $\pm$ 5	 & 	0.7 $\pm$ 0.1	 & 	57 $\pm$ 5	 & 	0.6 $\pm$ 0.1	 & 	51 $\pm$ 3	 & 	0.40 $\pm$ 0.01	 & 	1.06 $\pm$ 0.04	 & 	2.25 $\pm$ 0.13 \\ 
5	 & 	234.759399	 & 	-7.262989	 & 	2.7 $\pm$ 0.4	 & 	23 $\pm$ 4	 & 	1.6 $\pm$ 0.1	 & 	28 $\pm$ 2	 & 	1.7 $\pm$ 0.2	 & 	31 $\pm$ 4	 & 	0.35 $\pm$ 0.07	 & 	3.33 $\pm$ 1.15	 & 	1.95 $\pm$ 0.40 \\ 
\hline
\multicolumn{12}{c}{\bf MBM 34} \\
6	 & 	237.439326	 & 	-5.811189	 & 	1.2 $\pm$ 0.2	 & 	85 $\pm$ 4	 & 	1.1 $\pm$ 0.1	 & 	82 $\pm$ 2	 & 	0.7 $\pm$ 0.1	 & 	80 $\pm$ 3	 & 	0.48 $\pm$ 0.24	 & 	1.28 $\pm$ 0.65	 & 	2.67 $\pm$ 1.34 \\ 
7	 & 	237.521911	 & 	-5.932307	 & 	1.6 $\pm$ 0.2	 & 	79 $\pm$ 3	 & 	1.2 $\pm$ 0.1	 & 	80 $\pm$ 2	 & 	1.1 $\pm$ 0.1	 & 	84 $\pm$ 2	 & 	0.45 $\pm$ 0.01	 & 	1.67 $\pm$ 0.04	 & 	2.52 $\pm$ 0.15 \\ 
8	 & 	237.569725	 & 	-5.859141	 & 	1.3 $\pm$ 0.2	 & 	85 $\pm$ 4	 & 	1.1 $\pm$ 0.1	 & 	80 $\pm$ 2	 & 	1.0 $\pm$ 0.2	 & 	79 $\pm$ 3	 & 	0.52 $\pm$ 0.02	 & 	1.26 $\pm$ 0.03	 & 	2.91 $\pm$ 0.18 \\ 
9	 & 	237.417944	 & 	-5.917022	 & 	1.6 $\pm$ 0.3	 & 	81 $\pm$ 5	 & 	1.6 $\pm$ 0.1	 & 	79 $\pm$ 2	 & 	1.6 $\pm$ 0.1	 & 	80 $\pm$ 2	 & 	0.63 $\pm$ 0.02	 & 	1.69 $\pm$ 0.03	 & 	3.51 $\pm$ 0.22 \\ 
10	 & 	237.405578	 & 	-5.779075	 & 	1.6 $\pm$ 0.2	 & 	92 $\pm$ 3	 & 	1.5 $\pm$ 0.1	 & 	92 $\pm$ 1	 & 	1.4 $\pm$ 0.2	 &	92 $\pm$ 2	 & 	0.60 $\pm$ 0.05	 & 	1.61 $\pm$ 0.07	 & 	3.37 $\pm$ 0.32 \\ 
11	 & 	237.660440	 & 	-5.898259	 & 	1.7 $\pm$ 0.1	 & 	87 $\pm$ 2	 & 	1.4 $\pm$ 0.1	 & 	86 $\pm$ 1	 & 	1.4 $\pm$ 0.1	 & 	86 $\pm$ 1	 & 	0.54 $\pm$ 0.01	 & 	1.66 $\pm$ 0.02	 & 	3.04 $\pm$ 0.17 \\ 
12	 & 	237.287787	 & 	-5.928377	 & 	2.1 $\pm$ 0.1	 & 	98 $\pm$ 1	 & 	2.1 $\pm$ 0.1	 & 	98 $\pm$ 1	 & 	2.2 $\pm$ 0.1	 &	101 $\pm$ 1	 & 	0.68 $\pm$ 0.08	 & 	2.16 $\pm$ 0.09	 & 	3.80 $\pm$ 0.48 \\ 
13	 & 	237.674831	 & 	-5.749450	 & 	1.8 $\pm$ 0.2	 & 	86 $\pm$ 3	 & 	1.2 $\pm$ 0.2	 & 	81 $\pm$ 4	 & 	1.2 $\pm$ 0.3	 & 	80 $\pm$ 5	 & 	0.37 $\pm$ 0.03	 & 	2.21 $\pm$ 0.21	 & 	2.06 $\pm$ 0.18 \\ 
14	 & 	237.256837	 & 	-5.757691	 & 	1.7 $\pm$ 0.3	 & 	84 $\pm$ 4	 & 	1.4 $\pm$ 0.1	 & 	83 $\pm$ 1	 & 	1.3 $\pm$ 0.1	 & 	83 $\pm$ 2	 & 	0.48 $\pm$ 0.03	 & 	1.76 $\pm$ 0.13	 & 	2.69 $\pm$ 0.24 \\ 
15	 & 	237.745219	 & 	-5.842658	 & 	1.7 $\pm$ 0.3	 & 	87 $\pm$ 4	 & 	1.6 $\pm$ 0.1	 & 	84 $\pm$ 2	 & 	1.5 $\pm$ 0.2	 & 	85 $\pm$ 3	 & 	0.56 $\pm$ 0.04	 & 	1.75 $\pm$ 0.08	 & 	3.11 $\pm$ 0.26 \\ 
16	 & 	237.676361	 & 	-6.036192	 & 	1.0 $\pm$ 0.1	 & 	72 $\pm$ 4	 & 	1.0 $\pm$ 0.1	 & 	75 $\pm$ 2	 & 	0.9 $\pm$ 0.1	 & 	76 $\pm$ 3	 & 	0.59 $\pm$ 0.03	 & 	1.01 $\pm$ 0.03	 & 	3.32 $\pm$ 0.25 \\ 
17	 & 	237.760799	 & 	-6.024821	 & 	1.6 $\pm$ 0.3	 & 	76 $\pm$ 5	 & 	1.6 $\pm$ 0.1	 & 	79 $\pm$ 2	 & 	1.3 $\pm$ 0.2	 & 	85 $\pm$ 3	 & 	0.56 $\pm$ 0.13	 & 	1.67 $\pm$ 0.26	 & 	3.16 $\pm$ 0.75 \\ 
18	 & 	237.620548	 & 	-6.144651	 & 	1.7 $\pm$ 0.1	 & 	88 $\pm$ 1	 & 	1.2 $\pm$ 0.1	 & 	78 $\pm$ 1	 & 	1.2 $\pm$ 0.1	 & 	77 $\pm$ 1	 & 	0.44 $\pm$ 0.02	 & 	1.76 $\pm$ 0.08	 & 	2.46 $\pm$ 0.17 \\ 
19	 & 	237.684774	 & 	-6.159001	 & 	1.7 $\pm$ 0.3	 & 	86 $\pm$ 5	 & 	1.4 $\pm$ 0.2	 & 	82 $\pm$ 3	 & 	1.4 $\pm$ 0.2	 & 	81 $\pm$ 4	 & 	0.53 $\pm$ 0.02	 & 	1.67 $\pm$ 0.04	 & 	2.96 $\pm$ 0.19 \\ 
20	 & 	237.894932	 & 	-5.935640	 & 	1.6 $\pm$ 0.2	 & 	88 $\pm$ 2	 & 	1.4 $\pm$ 0.1	 & 	82 $\pm$ 2	 & 	1.4 $\pm$ 0.1	 & 	82 $\pm$ 2	 & 	0.56 $\pm$ 0.02	 & 	1.59 $\pm$ 0.04	 & 	3.13 $\pm$ 0.21 \\ 
21	 & 	237.916806	 & 	-5.985692	 & 	1.2 $\pm$ 0.2	 & 	84 $\pm$ 3	 & 	1.2 $\pm$ 0.1	 & 	81 $\pm$ 1	 & 	1.2 $\pm$ 0.1	 & 	82 $\pm$ 2	 & 	0.62 $\pm$ 0.00	 & 	1.26 $\pm$ 0.00	 & 	3.50 $\pm$ 0.19 \\ 
22	 & 	237.858470	 & 	-6.120708	 & 	1.3 $\pm$ 0.5	 & 	78 $\pm$ 1	 & 	1.2 $\pm$ 0.3	 & 	76 $\pm$ 5	 & 	1.2 $\pm$ 0.4	 & 	82 $\pm$ 8	 & 	0.63 $\pm$ 0.01	 & 	1.30 $\pm$ 0.01	 & 	3.50 $\pm$ 0.20 \\ 
23	 & 	237.746356	 & 	-6.276022	 & 	1.7 $\pm$ 0.4	 & 	76 $\pm$ 5	 & 	1.7 $\pm$ 0.3	 & 	82 $\pm$ 5	 & 	1.4 $\pm$ 0.3	 & 	83 $\pm$ 6	 & 	0.56 $\pm$ 0.08	 & 	1.72 $\pm$ 0.14	 & 	3.15 $\pm$ 0.45 \\ 
24	 & 	237.738667	 & 	-6.328300	 & 	1.4 $\pm$ 0.2	 & 	81 $\pm$ 3	 & 	1.4 $\pm$ 0.2	 & 	82 $\pm$ 4	 & 	1.2 $\pm$ 0.2	 & 	82 $\pm$ 3	 & 	0.59 $\pm$ 0.03	 & 	1.41 $\pm$ 0.04	 & 	3.32 $\pm$ 0.24 \\ 
25	 & 	237.979203	 & 	-6.117325	 & 	1.4 $\pm$ 0.1	 & 	81 $\pm$ 1	 & 	1.2 $\pm$ 0.1	 & 	77 $\pm$ 3	 & 	1.2 $\pm$ 0.1	 & 	81 $\pm$ 2	 & 	0.54 $\pm$ 0.02	 & 	1.45 $\pm$ 0.04	 & 	3.03 $\pm$ 0.21 \\ 
26	 & 	237.764368	 & 	-6.413710	 & 	1.2 $\pm$ 0.2	 & 	85 $\pm$ 4	 & 	1.1 $\pm$ 0.2	 & 	82 $\pm$ 4	 & 	1.1 $\pm$ 0.2	 & 	78 $\pm$ 5	 & 	0.58 $\pm$ 0.02	 & 	1.21 $\pm$ 0.02	 & 	3.26 $\pm$ 0.21 \\ 
\hline
\multicolumn{12}{c}{\bf MBM 35} \\
27	 & 	237.756650	 & 	-2.094230	 & 	1.6 $\pm$ 0.1	 & 	80 $\pm$ 2	 & 	1.5 $\pm$ 0.1	 & 	85 $\pm$ 1	 & 	1.4 $\pm$ 0.1	 & 	90 $\pm$ 2	 & 	0.58 $\pm$ 0.01	 & 	1.60 $\pm$ 0.02	 & 	3.24 $\pm$ 0.19 \\ 
28	 & 	237.790670	 & 	-2.142240	 & 	2.2 $\pm$ 0.2	 & 	78 $\pm$ 3	 & 	2.1 $\pm$ 0.1	 & 	82 $\pm$ 1	 & 	1.9 $\pm$ 0.2	 & 	87 $\pm$ 2	 & 	0.56 $\pm$ 0.03	 & 	2.26 $\pm$ 0.07	 & 	3.15 $\pm$ 0.22 \\ 
29	 & 	237.808010	 & 	-2.153560	 & 	2.0 $\pm$ 0.1	 & 	78 $\pm$ 2	 & 	1.9 $\pm$ 0.0	 & 	79 $\pm$ 1	 & 	1.7 $\pm$ 0.1	 & 	85 $\pm$ 1	 & 	0.55 $\pm$ 0.06	 & 	2.08 $\pm$ 0.17	 & 	3.06 $\pm$ 0.37 \\ 
30	 & 	237.830030	 & 	-2.148890	 & 	1.7 $\pm$ 0.5	 & 	75 $\pm$ 7	 & 	1.6 $\pm$ 0.3	 & 	81 $\pm$ 4	 & 	1.6 $\pm$ 0.3	 & 	90 $\pm$ 5	 & 	0.60 $\pm$ 0.01	 & 	1.72 $\pm$ 0.02	 & 	3.36 $\pm$ 0.19 \\ 
\hline
\end{tabular}
\end{sidewaystable}

\begin{table*}                                                                   
\begin{center}                                                                     
\caption{Polarization results of 234 stars observed in the direction of MBM33-39 cloud complex with P/$\sigma_{P}$ $\ge$ 2.}\label{ch7:Polresult}
\begin{tabular}{ccccccrr}\hline                                                       
Star & $\alpha$ (J2000)  & $\delta$ (J2000) & $l$ & $b$ & P $\pm$ $\epsilon_P$ & $\theta$ $\pm$ $\epsilon_{\theta}$  & $\theta_{Galactic}$\\ 
 ID  &   ($\degree$) &    ($\degree$)&  ($\degree$) &    ($\degree$)&  (\%)           & ($\degree$) & ($\degree$) \\\hline 
1	 & 	234.664825	 & 	-7.243520	 & 	358.732	 & 	36.915	 & 	1.2 $\pm$ 0.3	 & 	 77 $\pm$ 7	 & 125 \\ 
2	 & 	234.666275	 & 	-7.290588	 & 	358.690	 & 	36.882	 & 	0.7 $\pm$ 0.2	 & 	 78 $\pm$ 6	 & 126 \\ 
3	 & 	234.689941	 & 	-7.249604	 & 	358.747	 & 	36.892	 & 	0.7 $\pm$ 0.2	 & 	 80 $\pm$ 6	 & 128 \\ 
4	 & 	234.692627	 & 	-7.273990	 & 	358.727	 & 	36.874	 & 	0.6 $\pm$ 0.1	 & 	 72 $\pm$ 6	 & 120 \\ 
5	 & 	234.700455	 & 	-7.262742	 & 	358.744	 & 	36.875	 & 	1.5 $\pm$ 0.5	 & 	 60 $\pm$ 8	 & 108 \\ 
6	 & 	234.707031	 & 	-7.170176	 & 	358.835	 & 	36.932	 & 	1.7 $\pm$ 0.2	 & 	 66 $\pm$ 4	 & 114 \\ 
7	 & 	234.712326	 & 	-7.138198	 & 	358.870	 & 	36.950	 & 	0.6 $\pm$ 0.1	 & 	 70 $\pm$ 3	 & 118 \\ 
8	 & 	234.716324	 & 	-7.120708	 & 	358.889	 & 	36.959	 & 	1.0 $\pm$ 0.3	 & 	 67 $\pm$ 7	 & 115 \\ 
9	 & 	234.717743	 & 	-7.104168	 & 	358.906	 & 	36.969	 & 	0.7 $\pm$ 0.3	 & 	 57 $\pm$ 9	 & 105 \\ 
10	 & 	234.732742	 & 	-7.230433	 & 	358.801	 & 	36.873	 & 	0.4 $\pm$ 0.1	 & 	 71 $\pm$ 3	 & 119 \\ 
\multicolumn{8}{c}{}\\
11	 & 	234.736481	 & 	-7.179309	 & 	358.851	 & 	36.905	 & 	0.6 $\pm$ 0.1	 & 	 81 $\pm$ 3	 & 129 \\ 
12	 & 	234.739532	 & 	-7.227524	 & 	358.809	 & 	36.870	 & 	1.0 $\pm$ 0.5	 & 	 48 $\pm$ 11	 & 96 \\ 
13	 & 	234.745026	 & 	-7.086656	 & 	358.945	 & 	36.960	 & 	0.8 $\pm$ 0.1	 & 	 74 $\pm$ 5	 & 122 \\ 
14	 & 	234.747818	 & 	-7.236627	 & 	358.808	 & 	36.858	 & 	0.7 $\pm$ 0.1	 & 	 57 $\pm$ 5	 & 105 \\ 
15	 & 	234.760971	 & 	-7.097479	 & 	358.948	 & 	36.941	 & 	0.9 $\pm$ 0.2	 & 	 61 $\pm$ 7	 & 109 \\ 
16	 & 	234.824615	 & 	-7.261821	 & 	358.848	 & 	36.785	 & 	1.6 $\pm$ 0.4	 & 	 44 $\pm$ 7	 & 92 \\ 
17	 & 	234.890594	 & 	-7.439786	 & 	358.738	 & 	36.617	 & 	0.9 $\pm$ 0.3	 & 	 100 $\pm$ 8	 & 148 \\ 
18	 & 	234.896805	 & 	-7.027812	 & 	359.125	 & 	36.888	 & 	0.9 $\pm$ 0.1	 & 	 63 $\pm$ 2	 & 111 \\ 
19	 & 	234.925552	 & 	-7.420925	 & 	358.784	 & 	36.604	 & 	0.5 $\pm$ 0.2	 & 	 100 $\pm$ 11	 & 148 \\ 
20	 & 	234.930389	 & 	-7.418399	 & 	358.790	 & 	36.602	 & 	0.3 $\pm$ 0.1	 & 	 75 $\pm$ 6	 & 123 \\ 
\multicolumn{8}{c}{}\\
21	 & 	234.940918	 & 	-7.048465	 & 	359.142	 & 	36.841	 & 	1.9 $\pm$ 0.1	 & 	 78 $\pm$ 1	 & 126 \\ 
22	 & 	234.947113	 & 	-6.995152	 & 	359.197	 & 	36.872	 & 	2.4 $\pm$ 0.6	 & 	 65 $\pm$ 7	 & 113 \\ 
23	 & 	234.953140	 & 	-7.030041	 & 	359.170	 & 	36.844	 & 	2.3 $\pm$ 0.4	 & 	 77 $\pm$ 5	 & 125 \\ 
24	 & 	234.954727	 & 	-7.400796	 & 	358.827	 & 	36.596	 & 	1.5 $\pm$ 0.2	 & 	 89 $\pm$ 3	 & 137 \\ 
25	 & 	234.957733	 & 	-7.381444	 & 	358.847	 & 	36.606	 & 	1.3 $\pm$ 0.3	 & 	 83 $\pm$ 7	 & 131 \\ 
26	 & 	234.959579	 & 	-7.380103	 & 	358.850	 & 	36.606	 & 	1.2 $\pm$ 0.1	 & 	 73 $\pm$ 2	 & 121 \\ 
27	 & 	234.964523	 & 	-7.052097	 & 	359.159	 & 	36.821	 & 	2.6 $\pm$ 1.1	 & 	 93 $\pm$ 12	 & 141 \\ 
28	 & 	235.066238	 & 	-6.975558	 & 	359.314	 & 	36.797	 & 	2.2 $\pm$ 0.6	 & 	 96 $\pm$ 8	 & 144 \\ 
29	 & 	235.073257	 & 	-6.928510	 & 	359.363	 & 	36.823	 & 	1.1 $\pm$ 0.4	 & 	 84 $\pm$ 9	 & 132 \\ 
30	 & 	235.091385	 & 	-6.909038	 & 	359.397	 & 	36.823	 & 	2.7 $\pm$ 0.5	 & 	 80 $\pm$ 5	 & 128 \\ 
\multicolumn{8}{c}{}\\
31	 & 	235.096268	 & 	-7.356310	 & 	358.985	 & 	36.521	 & 	1.5 $\pm$ 0.1	 & 	 74 $\pm$ 2	 & 122 \\ 
32	 & 	235.102219	 & 	-7.320977	 & 	359.023	 & 	36.540	 & 	2.1 $\pm$ 0.6	 & 	 90 $\pm$ 20	 & 138 \\ 
33	 & 	235.102951	 & 	-7.296313	 & 	359.046	 & 	36.556	 & 	1.0 $\pm$ 0.2	 & 	 75 $\pm$ 5	 & 123 \\ 
34	 & 	235.104874	 & 	-7.352910	 & 	358.995	 & 	36.517	 & 	1.0 $\pm$ 0.5	 & 	 82 $\pm$ 16	 & 130 \\ 
35	 & 	235.154037	 & 	-7.247395	 & 	359.133	 & 	36.551	 & 	1.6 $\pm$ 0.5	 & 	 89 $\pm$ 9	 & 137 \\ 
36	 & 	235.163086	 & 	-7.247187	 & 	359.141	 & 	36.544	 & 	1.8 $\pm$ 0.1	 & 	 91 $\pm$ 1	 & 139 \\ 
37	 & 	235.166168	 & 	-6.948177	 & 	359.422	 & 	36.741	 & 	1.6 $\pm$ 0.1	 & 	 78 $\pm$ 1	 & 126 \\ 
38	 & 	235.167191	 & 	-6.953535	 & 	359.418	 & 	36.737	 & 	2.3 $\pm$ 0.7	 & 	 89 $\pm$ 9	 & 137 \\ 
39	 & 	235.182129	 & 	-7.263934	 & 	359.141	 & 	36.519	 & 	1.4 $\pm$ 0.3	 & 	 117 $\pm$ 6	 & 165 \\ 
40	 & 	235.185669	 & 	-7.191573	 & 	359.211	 & 	36.565	 & 	2.0 $\pm$ 0.1	 & 	 100 $\pm$ 2	 & 148 \\ 
\multicolumn{8}{c}{}\\
41	 & 	235.206650	 & 	-7.242327	 & 	359.181	 & 	36.515	 & 	2.2 $\pm$ 0.1	 & 	 85 $\pm$ 1	 & 133 \\ 
42	 & 	235.243393	 & 	-6.988899	 & 	359.447	 & 	36.657	 & 	1.0 $\pm$ 0.1	 & 	 96 $\pm$ 4	 & 144 \\ 
43	 & 	235.253021	 & 	-6.980438	 & 	359.463	 & 	36.655	 & 	1.4 $\pm$ 0.1	 & 	 97 $\pm$ 2	 & 145 \\ 
44	 & 	235.306229	 & 	-7.133462	 & 	359.364	 & 	36.514	 & 	1.4 $\pm$ 0.5	 & 	 89 $\pm$ 9	 & 137 \\ 
45	 & 	235.322968	 & 	-7.150585	 & 	359.362	 & 	36.490	 & 	1.2 $\pm$ 0.1	 & 	 78 $\pm$ 2	 & 126 \\ 
46	 & 	235.329102	 & 	-7.145841	 & 	359.372	 & 	36.489	 & 	1.6 $\pm$ 0.5	 & 	 69 $\pm$ 8	 & 117 \\ 
47	 & 	235.349609	 & 	-7.066414	 & 	359.462	 & 	36.526	 & 	1.3 $\pm$ 0.2	 & 	 102 $\pm$ 4	 & 150 \\ 
48	 & 	235.367905	 & 	-7.076525	 & 	359.468	 & 	36.506	 & 	1.7 $\pm$ 0.3	 & 	 66 $\pm$ 5	 & 114 \\ 
49	 & 	235.379730	 & 	-7.104511	 & 	359.451	 & 	36.479	 & 	0.8 $\pm$ 0.2	 & 	 90 $\pm$ 6	 & 138 \\ 
50	 & 	237.253870	 & 	-1.864780	 & 	6.039	 & 	38.372	 & 	1.4 $\pm$ 0.1	 & 	 92 $\pm$ 2	 & 144 \\ 
\multicolumn{8}{c}{}\\
51	 & 	237.255371	 & 	-5.878733	 & 	2.108	 & 	35.864	 & 	1.6 $\pm$ 0.6	 & 	 63 $\pm$ 10	 & 113 \\ 
52	 & 	237.255870	 & 	-1.874110	 & 	6.032	 & 	38.364	 & 	1.5 $\pm$ 0.3	 & 	 87 $\pm$ 5	 & 140 \\ 
53	 & 	237.265854	 & 	-5.902115	 & 	2.094	 & 	35.842	 & 	1.5 $\pm$ 0.2	 & 	 76 $\pm$ 5	 & 126 \\ 
54	 & 	237.275220	 & 	-1.878780	 & 	6.042	 & 	38.345	 & 	1.7 $\pm$ 0.1	 & 	 82 $\pm$ 2	 & 134 \\ 
55	 & 	237.279709	 & 	-5.894644	 & 	2.112	 & 	35.836	 & 	2.4 $\pm$ 0.5	 & 	 72 $\pm$ 6	 & 122 \\ 
56	 & 	237.280550	 & 	-1.853440	 & 	6.071	 & 	38.357	 & 	1.6 $\pm$ 0.3	 & 	 87 $\pm$ 5	 & 140 \\ 
57	 & 	237.289220	 & 	-1.902780	 & 	6.029	 & 	38.320	 & 	1.2 $\pm$ 0.3	 & 	 89 $\pm$ 6	 & 142 \\ 
58	 & 	237.298798	 & 	-5.929773	 & 	2.094	 & 	35.799	 & 	1.1 $\pm$ 0.1	 & 	 87 $\pm$ 2	 & 137 \\ 
59	 & 	237.303900	 & 	-1.938780	 & 	6.004	 & 	38.286	 & 	1.1 $\pm$ 0.3	 & 	 78 $\pm$ 8	 & 131 \\ 
60	 & 	237.308578	 & 	-5.884315	 & 	2.144	 & 	35.820	 & 	1.4 $\pm$ 0.3	 & 	 79 $\pm$ 6	 & 129 \\ \hline
\end{tabular}
\end{center}  
\end{table*} 

\begin{table*}                                                                   
\begin{center}                                                                     
\contcaption{}
\begin{tabular}{ccccccrr}\hline                                                       
Star & $\alpha$ (J2000)  & $\delta$ (J2000)  & $l$ & $b$ & P $\pm$ $\epsilon_P$ & $\theta$ $\pm$ $\epsilon_{\theta}$  & $\theta_{Galactic}$\\ 
 ID  &   ($\degree$) &    ($\degree$)&  ($\degree$) &    ($\degree$)&  (\%)           & ($\degree$) & ($\degree$) \\\hline   
61	 & 	237.311900	 & 	-1.891440	 & 	6.058	 & 	38.309	 & 	1.5 $\pm$ 0.1	 & 	 83 $\pm$ 1	 & 136 \\ 
62	 & 	237.320938	 & 	-5.921296	 & 	2.119	 & 	35.787	 & 	1.9 $\pm$ 0.3	 & 	 87 $\pm$ 5	 & 137 \\ 
63	 & 	237.330383	 & 	-5.922173	 & 	2.126	 & 	35.779	 & 	1.4 $\pm$ 0.2	 & 	 84 $\pm$ 4	 & 134 \\ 
64	 & 	237.335815	 & 	-5.898364	 & 	2.152	 & 	35.790	 & 	1.5 $\pm$ 0.3	 & 	 93 $\pm$ 5	 & 143 \\ 
65	 & 	237.344498	 & 	-5.892202	 & 	2.165	 & 	35.788	 & 	1.3 $\pm$ 0.4	 & 	 87 $\pm$ 9	 & 137 \\ 
66	 & 	237.351700	 & 	-5.949665	 & 	2.116	 & 	35.745	 & 	1.1 $\pm$ 0.1	 & 	 78 $\pm$ 4	 & 128 \\ 
67	 & 	237.351868	 & 	-5.780048	 & 	2.277	 & 	35.854	 & 	1.0 $\pm$ 0.2	 & 	 72 $\pm$ 5	 & 122 \\ 
68	 & 	237.352875	 & 	-5.859034	 & 	2.203	 & 	35.803	 & 	1.4 $\pm$ 0.1	 & 	 76 $\pm$ 2	 & 126 \\ 
69	 & 	237.352890	 & 	-5.858979	 & 	2.203	 & 	35.803	 & 	1.2 $\pm$ 0.1	 & 	 85 $\pm$ 2	 & 135 \\ 
70	 & 	237.353088	 & 	-5.775971	 & 	2.282	 & 	35.855	 & 	1.9 $\pm$ 0.3	 & 	 87 $\pm$ 5	 & 137 \\ 
\multicolumn{8}{c}{}\\
71	 & 	237.365494	 & 	-5.903028	 & 	2.171	 & 	35.765	 & 	2.0 $\pm$ 0.6	 & 	 75 $\pm$ 8	 & 125 \\ 
72	 & 	237.374619	 & 	-5.854682	 & 	2.224	 & 	35.789	 & 	1.2 $\pm$ 0.5	 & 	 90 $\pm$ 12	 & 140 \\ 
73	 & 	237.391388	 & 	-5.759244	 & 	2.328	 & 	35.837	 & 	1.9 $\pm$ 0.5	 & 	 83 $\pm$ 8	 & 133 \\ 
74	 & 	237.404816	 & 	-6.003438	 & 	2.107	 & 	35.670	 & 	1.3 $\pm$ 0.2	 & 	 81 $\pm$ 4	 & 131 \\ 
75	 & 	237.405304	 & 	-5.904298	 & 	2.201	 & 	35.733	 & 	1.8 $\pm$ 0.5	 & 	 74 $\pm$ 7	 & 124 \\ 
76	 & 	237.408081	 & 	-5.839082	 & 	2.265	 & 	35.773	 & 	1.6 $\pm$ 0.2	 & 	 82 $\pm$ 4	 & 132 \\ 
77	 & 	237.417755	 & 	-5.992552	 & 	2.127	 & 	35.667	 & 	1.7 $\pm$ 0.6	 & 	 67 $\pm$ 9	 & 117 \\ 
78	 & 	237.424026	 & 	-5.955384	 & 	2.168	 & 	35.687	 & 	1.7 $\pm$ 0.3	 & 	 89 $\pm$ 4	 & 139 \\ 
79	 & 	237.426605	 & 	-5.901800	 & 	2.220	 & 	35.719	 & 	1.8 $\pm$ 0.5	 & 	 64 $\pm$ 8	 & 114 \\ 
80	 & 	237.437469	 & 	-5.903768	 & 	2.227	 & 	35.709	 & 	3.0 $\pm$ 0.6	 & 	 83 $\pm$ 6	 & 133 \\ 
\multicolumn{8}{c}{}\\
81	 & 	237.437714	 & 	-5.785320	 & 	2.340	 & 	35.785	 & 	2.2 $\pm$ 0.6	 & 	 87 $\pm$ 8	 & 137 \\ 
82	 & 	237.448563	 & 	-5.951352	 & 	2.191	 & 	35.670	 & 	1.7 $\pm$ 0.2	 & 	 88 $\pm$ 4	 & 138 \\ 
83	 & 	237.448563	 & 	-5.985734	 & 	2.158	 & 	35.648	 & 	1.8 $\pm$ 0.6	 & 	 95 $\pm$ 9	 & 145 \\ 
84	 & 	237.452255	 & 	-6.092173	 & 	2.060	 & 	35.577	 & 	1.6 $\pm$ 0.1	 & 	 84 $\pm$ 3	 & 134 \\ 
85	 & 	237.487640	 & 	-6.077678	 & 	2.102	 & 	35.560	 & 	1.7 $\pm$ 0.4	 & 	 78 $\pm$ 7	 & 128 \\ 
86	 & 	237.493164	 & 	-6.100979	 & 	2.084	 & 	35.541	 & 	1.6 $\pm$ 0.6	 & 	 74 $\pm$ 10	 & 124 \\ 
87	 & 	237.501923	 & 	-6.037074	 & 	2.151	 & 	35.575	 & 	1.5 $\pm$ 0.2	 & 	 72 $\pm$ 4	 & 122 \\ 
88	 & 	237.537659	 & 	-5.914225	 & 	2.295	 & 	35.626	 & 	3.2 $\pm$ 0.9	 & 	 73 $\pm$ 8	 & 123 \\ 
89	 & 	237.547300	 & 	-1.594130	 & 	6.541	 & 	38.302	 & 	0.8 $\pm$ 0.1	 & 	 81 $\pm$ 2	 & 134 \\ 
90	 & 	237.556381	 & 	-5.910893	 & 	2.313	 & 	35.614	 & 	2.1 $\pm$ 0.6	 & 	 90 $\pm$ 8	 & 140 \\ 
\multicolumn{8}{c}{}\\
91	 & 	237.561127	 & 	-5.861592	 & 	2.364	 & 	35.641	 & 	1.3 $\pm$ 0.5	 & 	 102 $\pm$ 10	 & 152 \\ 
92	 & 	237.573608	 & 	-5.892675	 & 	2.344	 & 	35.612	 & 	1.4 $\pm$ 0.4	 & 	 103 $\pm$ 8	 & 153 \\ 
93	 & 	237.577881	 & 	-5.833573	 & 	2.403	 & 	35.646	 & 	1.4 $\pm$ 0.2	 & 	 85 $\pm$ 5	 & 135 \\ 
94	 & 	237.585510	 & 	-5.881410	 & 	2.364	 & 	35.610	 & 	2.4 $\pm$ 0.6	 & 	 88 $\pm$ 7	 & 138 \\ 
95	 & 	237.606903	 & 	-6.225999	 & 	2.055	 & 	35.374	 & 	1.9 $\pm$ 0.3	 & 	 85 $\pm$ 4	 & 135 \\ 
96	 & 	237.608292	 & 	-5.869369	 & 	2.393	 & 	35.600	 & 	1.6 $\pm$ 0.2	 & 	 85 $\pm$ 4	 & 135 \\ 
97	 & 	237.616135	 & 	-6.208387	 & 	2.079	 & 	35.378	 & 	1.8 $\pm$ 0.2	 & 	 77 $\pm$ 3	 & 127 \\ 
98	 & 	237.642258	 & 	-6.230644	 & 	2.078	 & 	35.344	 & 	1.6 $\pm$ 0.4	 & 	 78 $\pm$ 7	 & 128 \\ 
99	 & 	237.674896	 & 	-6.129722	 & 	2.199	 & 	35.383	 & 	1.7 $\pm$ 0.2	 & 	 87 $\pm$ 4	 & 137 \\ 
100	 & 	237.682358	 & 	-5.980763	 & 	2.345	 & 	35.473	 & 	1.9 $\pm$ 0.3	 & 	 81 $\pm$ 4	 & 131 \\ 
\multicolumn{8}{c}{}\\
101	 & 	237.688000	 & 	-1.452510	 & 	6.793	 & 	38.275	 & 	1.5 $\pm$ 0.3	 & 	 83 $\pm$ 5	 & 136 \\ 
102	 & 	237.704025	 & 	-5.950323	 & 	2.391	 & 	35.475	 & 	1.5 $\pm$ 0.3	 & 	 74 $\pm$ 6	 & 124 \\ 
103	 & 	237.709381	 & 	-6.003910	 & 	2.344	 & 	35.437	 & 	1.7 $\pm$ 0.6	 & 	 85 $\pm$ 10	 & 135 \\ 
104	 & 	237.711685	 & 	-6.041290	 & 	2.311	 & 	35.411	 & 	1.3 $\pm$ 0.1	 & 	 81 $\pm$ 2	 & 131 \\ 
105	 & 	237.719116	 & 	-6.180243	 & 	2.185	 & 	35.317	 & 	1.4 $\pm$ 0.1	 & 	 82 $\pm$ 3	 & 132 \\ 
106	 & 	237.720078	 & 	-5.966722	 & 	2.388	 & 	35.453	 & 	1.3 $\pm$ 0.3	 & 	 85 $\pm$ 5	 & 135 \\ 
107	 & 	237.721649	 & 	-5.970223	 & 	2.386	 & 	35.449	 & 	1.2 $\pm$ 0.4	 & 	 94 $\pm$ 9	 & 144 \\ 
108	 & 	237.724228	 & 	-6.130367	 & 	2.237	 & 	35.345	 & 	1.4 $\pm$ 0.4	 & 	 91 $\pm$ 8	 & 141 \\ 
109	 & 	237.724548	 & 	-6.144137	 & 	2.224	 & 	35.336	 & 	0.7 $\pm$ 0.2	 & 	 77 $\pm$ 8	 & 127 \\ 
110	 & 	237.728363	 & 	-5.992858	 & 	2.370	 & 	35.430	 & 	1.4 $\pm$ 0.2	 & 	 81 $\pm$ 4	 & 131 \\ 
\multicolumn{8}{c}{}\\
111	 & 	237.730743	 & 	-6.148084	 & 	2.225	 & 	35.329	 & 	1.8 $\pm$ 0.4	 & 	 93 $\pm$ 6	 & 143 \\ 
112	 & 	237.734010	 & 	-1.375170	 & 	6.907	 & 	38.284	 & 	1.5 $\pm$ 0.3	 & 	 81 $\pm$ 5	 & 134 \\ 
113	 & 	237.736496	 & 	-6.106158	 & 	2.269	 & 	35.351	 & 	1.7 $\pm$ 0.3	 & 	 89 $\pm$ 3	 & 139 \\ 
114	 & 	237.736680	 & 	-1.432510	 & 	6.851	 & 	38.248	 & 	1.5 $\pm$ 0.7	 & 	 90 $\pm$ 13	 & 143 \\ 
115	 & 	237.737350	 & 	-1.415840	 & 	6.868	 & 	38.257	 & 	1.3 $\pm$ 0.3	 & 	 80 $\pm$ 6	 & 133 \\ 
116	 & 	237.739227	 & 	-6.099532	 & 	2.277	 & 	35.353	 & 	1.6 $\pm$ 0.6	 & 	 67 $\pm$ 10	 & 117 \\ 
117	 & 	237.741058	 & 	-6.186683	 & 	2.196	 & 	35.296	 & 	1.7 $\pm$ 0.4	 & 	 82 $\pm$ 7	 & 132 \\ 
118	 & 	237.756650	 & 	-2.094230	 & 	6.196	 & 	37.833	 & 	1.5 $\pm$ 0.1	 & 	 85 $\pm$ 1	 & 138 \\ 
119	 & 	237.760818	 & 	-6.024866	 & 	2.365	 & 	35.384	 & 	1.5 $\pm$ 0.3	 & 	 85 $\pm$ 4	 & 135 \\ 
120	 & 	237.768234	 & 	-6.146307	 & 	2.256	 & 	35.301	 & 	1.9 $\pm$ 0.8	 & 	 107 $\pm$ 12	 & 157 \\ \hline
\end{tabular}
\end{center}  
\end{table*} 

\begin{table*}                                                                   
\begin{center}                                                                     
\contcaption{}
\begin{tabular}{ccccccrr}\hline                                                       
Star & $\alpha$ (J2000)  & $\delta$ (J2000)  & $l$ & $b$ & P $\pm$ $\epsilon_P$ & $\theta$ $\pm$ $\epsilon_{\theta}$  & $\theta_{Galactic}$\\ 
 ID  &   ($\degree$) &    ($\degree$)&  ($\degree$) &    ($\degree$)&  (\%)           & ($\degree$) & ($\degree$) \\\hline   
121	 & 	237.785873	 & 	-6.090573	 & 	2.322	 & 	35.323	 & 	1.7 $\pm$ 0.5	 & 	 77 $\pm$ 6	 & 127 \\ 
122	 & 	237.790670	 & 	-2.142240	 & 	6.174	 & 	37.776	 & 	2.1 $\pm$ 0.1	 & 	 82 $\pm$ 1	 & 135 \\ 
123	 & 	237.808010	 & 	-2.153560	 & 	6.176	 & 	37.756	 & 	1.9 $\pm$ 0.1	 & 	 79 $\pm$ 1	 & 132 \\ 
124	 & 	237.830030	 & 	-2.148890	 & 	6.197	 & 	37.741	 & 	1.6 $\pm$ 0.3	 & 	 81 $\pm$ 4	 & 134 \\ 
125	 & 	237.850950	 & 	-1.513970	 & 	6.855	 & 	38.107	 & 	1.4 $\pm$ 0.4	 & 	 79 $\pm$ 8	 & 132 \\ 
126	 & 	237.861950	 & 	-1.347710	 & 	7.031	 & 	38.199	 & 	1.0 $\pm$ 0.2	 & 	 71 $\pm$ 4	 & 125 \\ 
127	 & 	237.865620	 & 	-1.513300	 & 	6.867	 & 	38.096	 & 	1.5 $\pm$ 0.1	 & 	 84 $\pm$ 2	 & 137 \\ 
128	 & 	237.886960	 & 	-1.545300	 & 	6.850	 & 	38.060	 & 	1.6 $\pm$ 0.3	 & 	 82 $\pm$ 5	 & 136 \\ 
129	 & 	237.895290	 & 	-1.364380	 & 	7.041	 & 	38.162	 & 	1.2 $\pm$ 0.2	 & 	 82 $\pm$ 5	 & 136 \\ 
130	 & 	237.960640	 & 	-1.363040	 & 	7.092	 & 	38.110	 & 	2.1 $\pm$ 0.1	 & 	 106 $\pm$ 1	 & 160 \\ 
\multicolumn{8}{c}{}\\
131	 & 	238.023588	 & 	-4.332006	 & 	4.189	 & 	36.248	 & 	5.3 $\pm$ 2.6	 & 	 66 $\pm$ 14	 & 118 \\ 
132	 & 	238.050981	 & 	-4.372278	 & 	4.171	 & 	36.201	 & 	3.6 $\pm$ 1.2	 & 	 69 $\pm$ 9	 & 120 \\ 
133	 & 	238.071911	 & 	-4.324807	 & 	4.233	 & 	36.214	 & 	2.4 $\pm$ 0.1	 & 	 74 $\pm$ 1	 & 126 \\ 
134	 & 	238.082026	 & 	-4.336071	 & 	4.230	 & 	36.200	 & 	5.1 $\pm$ 2.1	 & 	 73 $\pm$ 11	 & 124 \\ 
135	 & 	238.167678	 & 	-4.207979	 & 	4.420	 & 	36.212	 & 	1.6 $\pm$ 0.1	 & 	 65 $\pm$ 2	 & 116 \\ 
136	 & 	238.173347	 & 	-4.248406	 & 	4.385	 & 	36.183	 & 	2.1 $\pm$ 0.1	 & 	 72 $\pm$ 1	 & 124 \\ 
137	 & 	238.189188	 & 	-4.824758	 & 	3.840	 & 	35.811	 & 	1.9 $\pm$ 0.1	 & 	 64 $\pm$ 1	 & 115 \\ 
138	 & 	238.196004	 & 	-4.175669	 & 	4.473	 & 	36.210	 & 	1.9 $\pm$ 0.4	 & 	 78 $\pm$ 6	 & 130 \\ 
139	 & 	238.209161	 & 	-4.165420	 & 	4.494	 & 	36.206	 & 	2.1 $\pm$ 0.1	 & 	 76 $\pm$ 1	 & 128 \\ 
140	 & 	238.224603	 & 	-4.210609	 & 	4.461	 & 	36.166	 & 	1.9 $\pm$ 0.3	 & 	 75 $\pm$ 5	 & 127 \\ 
\multicolumn{8}{c}{}\\
141	 & 	238.229792	 & 	-4.870920	 & 	3.827	 & 	35.751	 & 	1.7 $\pm$ 0.1	 & 	 69 $\pm$ 1	 & 120 \\ 
142	 & 	238.233564	 & 	-4.158268	 & 	4.519	 & 	36.191	 & 	2.3 $\pm$ 0.2	 & 	 74 $\pm$ 2	 & 125 \\ 
143	 & 	238.267636	 & 	-4.250212	 & 	4.456	 & 	36.108	 & 	1.2 $\pm$ 0.4	 & 	 65 $\pm$ 8	 & 116 \\ 
144	 & 	238.275569	 & 	-4.203571	 & 	4.507	 & 	36.131	 & 	1.8 $\pm$ 0.1	 & 	 81 $\pm$ 2	 & 133 \\ 
145	 & 	238.314770	 & 	-3.194860	 & 	5.525	 & 	36.721	 & 	2.0 $\pm$ 0.1	 & 	 83 $\pm$ 1	 & 135 \\ 
146	 & 	238.341834	 & 	-4.378782	 & 	4.388	 & 	35.970	 & 	2.6 $\pm$ 0.1	 & 	 76 $\pm$ 1	 & 127 \\ 
147	 & 	238.355248	 & 	-4.347309	 & 	4.429	 & 	35.979	 & 	2.7 $\pm$ 0.1	 & 	 78 $\pm$ 1	 & 130 \\ 
148	 & 	238.365447	 & 	-4.299565	 & 	4.483	 & 	36.001	 & 	2.5 $\pm$ 0.2	 & 	 77 $\pm$ 2	 & 129 \\ 
149	 & 	238.449810	 & 	-3.034300	 & 	5.787	 & 	36.712	 & 	3.1 $\pm$ 0.1	 & 	 84 $\pm$ 1	 & 137 \\ 
150	 & 	238.493830	 & 	-3.042060	 & 	5.812	 & 	36.672	 & 	2.5 $\pm$ 0.1	 & 	 102 $\pm$ 1	 & 155 \\ 
\multicolumn{8}{c}{}\\
151	 & 	238.576830	 & 	-3.202710	 & 	5.717	 & 	36.509	 & 	3.6 $\pm$ 0.2	 & 	 94 $\pm$ 2	 & 146 \\ 
152	 & 	238.581580	 & 	-3.129170	 & 	5.793	 & 	36.550	 & 	1.8 $\pm$ 0.1	 & 	 95 $\pm$ 2	 & 148 \\ 
153	 & 	238.608153	 & 	-4.895051	 & 	4.093	 & 	35.441	 & 	2.5 $\pm$ 0.1	 & 	 97 $\pm$ 2	 & 148 \\ 
154	 & 	238.625497	 & 	-4.900464	 & 	4.101	 & 	35.424	 & 	1.7 $\pm$ 0.1	 & 	 89 $\pm$ 2	 & 140 \\ 
155	 & 	238.667280	 & 	-3.364380	 & 	5.626	 & 	36.339	 & 	3.1 $\pm$ 0.1	 & 	 96 $\pm$ 1	 & 148 \\ 
156	 & 	238.714760	 & 	-3.395930	 & 	5.631	 & 	36.282	 & 	3.4 $\pm$ 0.1	 & 	 105 $\pm$ 1	 & 157 \\ 
157	 & 	238.746060	 & 	-3.520360	 & 	5.532	 & 	36.181	 & 	2.3 $\pm$ 0.1	 & 	 94 $\pm$ 1	 & 146 \\ 
158	 & 	238.747170	 & 	-3.400650	 & 	5.651	 & 	36.253	 & 	2.8 $\pm$ 0.1	 & 	 100 $\pm$ 1	 & 152 \\ 
159	 & 	238.752880	 & 	-4.419318	 & 	4.662	 & 	35.623	 & 	1.8 $\pm$ 0.2	 & 	 89 $\pm$ 3	 & 141 \\ 
160	 & 	238.756890	 & 	-3.389780	 & 	5.668	 & 	36.252	 & 	2.2 $\pm$ 0.1	 & 	 87 $\pm$ 1	 & 139 \\ 
\multicolumn{8}{c}{}\\
161	 & 	238.785680	 & 	-3.513780	 & 	5.568	 & 	36.154	 & 	2.7 $\pm$ 0.1	 & 	 94 $\pm$ 2	 & 146 \\ 
162	 & 	238.795840	 & 	-3.177160	 & 	5.907	 & 	36.351	 & 	2.7 $\pm$ 0.1	 & 	 89 $\pm$ 1	 & 142 \\ 
163	 & 	238.804692	 & 	-4.371314	 & 	4.747	 & 	35.612	 & 	1.9 $\pm$ 0.3	 & 	 85 $\pm$ 5	 & 137 \\ 
164	 & 	238.808233	 & 	-4.401561	 & 	4.721	 & 	35.591	 & 	1.9 $\pm$ 0.2	 & 	 88 $\pm$ 2	 & 140 \\ 
165	 & 	238.813560	 & 	-3.165470	 & 	5.932	 & 	36.344	 & 	2.5 $\pm$ 0.1	 & 	 105 $\pm$ 1	 & 158 \\ 
166	 & 	238.871850	 & 	-3.287220	 & 	5.857	 & 	36.223	 & 	2.2 $\pm$ 0.1	 & 	 88 $\pm$ 1	 & 141 \\ 
167	 & 	238.893670	 & 	-3.314560	 & 	5.846	 & 	36.190	 & 	2.2 $\pm$ 0.1	 & 	 86 $\pm$ 1	 & 139 \\ 
168	 & 	238.900830	 & 	-3.313320	 & 	5.853	 & 	36.185	 & 	2.1 $\pm$ 0.1	 & 	 85 $\pm$ 1	 & 138 \\ 
169	 & 	238.908540	 & 	-3.146380	 & 	6.022	 & 	36.280	 & 	4.6 $\pm$ 0.3	 & 	 112 $\pm$ 2	 & 165 \\ 
170	 & 	238.923610	 & 	-3.184580	 & 	5.996	 & 	36.245	 & 	3.2 $\pm$ 0.1	 & 	 105 $\pm$ 1	 & 158 \\ 
\multicolumn{8}{c}{}\\
171	 & 	238.938600	 & 	-3.169830	 & 	6.022	 & 	36.242	 & 	2.8 $\pm$ 0.2	 & 	 101 $\pm$ 2	 & 154 \\ 
172	 & 	239.044520	 & 	-3.325680	 & 	5.948	 & 	36.063	 & 	2.4 $\pm$ 0.1	 & 	 92 $\pm$ 1	 & 145 \\ 
173	 & 	239.047040	 & 	-3.327360	 & 	5.948	 & 	36.060	 & 	2.6 $\pm$ 0.1	 & 	 90 $\pm$ 1	 & 143 \\ 
174	 & 	239.155790	 & 	-3.388940	 & 	5.969	 & 	35.937	 & 	2.6 $\pm$ 0.1	 & 	 98 $\pm$ 1	 & 151 \\ 
175	 & 	239.489430	 & 	-3.547650	 & 	6.063	 & 	35.576	 & 	2.0 $\pm$ 0.2	 & 	 88 $\pm$ 3	 & 141 \\ 
176	 & 	239.518830	 & 	-3.587160	 & 	6.046	 & 	35.528	 & 	2.1 $\pm$ 0.2	 & 	 84 $\pm$ 3	 & 137 \\ 
177	 & 	239.521390	 & 	-3.574160	 & 	6.061	 & 	35.534	 & 	2.0 $\pm$ 0.1	 & 	 92 $\pm$ 2	 & 145 \\ 
178	 & 	239.521460	 & 	-3.542270	 & 	6.092	 & 	35.553	 & 	2.0 $\pm$ 0.1	 & 	 86 $\pm$ 2	 & 139 \\ 
179	 & 	239.799280	 & 	-1.526560	 & 	8.298	 & 	36.533	 & 	1.2 $\pm$ 0.3	 & 	 80 $\pm$ 6	 & 134 \\ 
180	 & 	239.803790	 & 	-1.930900	 & 	7.894	 & 	36.291	 & 	1.6 $\pm$ 0.2	 & 	 89 $\pm$ 3	 & 143 \\ \hline
\end{tabular}
\end{center}  
\end{table*} 

\begin{table*}                                                           

\begin{center}                                                                     
\contcaption{}
\begin{threeparttable} 
\begin{tabular}{ccccccrr}\hline                                                       
Star & $\alpha$ (J2000)  & $\delta$ (J2000)  & $l$ & $b$ & P $\pm$ $\epsilon_P$ & $\theta$ $\pm$ $\epsilon_{\theta}$  & $\theta_{Galactic}$\\ 
 ID  &   ($\degree$) &    ($\degree$)&  ($\degree$) &    ($\degree$)&  (\%)           & ($\degree$) & ($\degree$) \\\hline   
181	 & 	239.805130	 & 	-1.875570	 & 	7.951	 & 	36.323	 & 	1.3 $\pm$ 0.3	 & 	 103 $\pm$ 6	 & 157 \\ 
182	 & 	239.821140	 & 	-1.919570	 & 	7.919	 & 	36.284	 & 	1.3 $\pm$ 0.1	 & 	 87 $\pm$ 2	 & 140 \\ 
183	 & 	239.833960	 & 	-1.483890	 & 	8.366	 & 	36.530	 & 	0.6 $\pm$ 0.3	 & 	 88 $\pm$ 10	 & 142 \\ 
184	 & 	239.851300	 & 	-1.504560	 & 	8.357	 & 	36.504	 & 	0.5 $\pm$ 0.1	 & 	 88 $\pm$ 6	 & 142 \\ 
185	 & 	239.855970	 & 	-1.506560	 & 	8.358	 & 	36.499	 & 	0.7 $\pm$ 0.1	 & 	 86 $\pm$ 4	 & 140 \\ 
186	 & 	239.862640	 & 	-1.432560	 & 	8.438	 & 	36.538	 & 	0.5 $\pm$ 0.2	 & 	 65 $\pm$ 9	 & 119 \\ 
187	 & 	239.864640	 & 	-1.466560	 & 	8.405	 & 	36.515	 & 	0.6 $\pm$ 0.3	 & 	 60 $\pm$ 11	 & 114 \\ 
188	 & 	239.899280	 & 	-1.749280	 & 	8.146	 & 	36.322	 & 	1.9 $\pm$ 0.5	 & 	 89 $\pm$ 8	 & 143 \\ 
189	 & 	239.906260	 & 	-1.302680	 & 	8.601	 & 	36.578	 & 	0.8 $\pm$ 0.1	 & 	 77 $\pm$ 3	 & 132 \\ 
190	 & 	239.918250	 & 	-1.800980	 & 	8.108	 & 	36.276	 & 	1.7 $\pm$ 0.1	 & 	 95 $\pm$ 1	 & 149 \\ 
\multicolumn{8}{c}{}\\
191	 & 	239.953610	 & 	-1.328680	 & 	8.609	 & 	36.525	 & 	0.9 $\pm$ 0.2	 & 	 76 $\pm$ 5	 & 130 \\ 
192	 & 	239.966280	 & 	-1.316080	 & 	8.636	 & 	36.525	 & 	1.0 $\pm$ 0.2	 & 	 75 $\pm$ 5	 & 130 \\ 
193	 & 	239.974950	 & 	-1.316070	 & 	8.635	 & 	36.519	 & 	1.8 $\pm$ 0.4	 & 	 81 $\pm$ 6	 & 135 \\ 
194	 & 	239.983780	 & 	-1.743370	 & 	8.214	 & 	36.256	 & 	2.1 $\pm$ 0.1	 & 	 97 $\pm$ 1	 & 151 \\ 
195	 & 	240.067820	 & 	-1.710040	 & 	8.309	 & 	36.209	 & 	1.7 $\pm$ 0.2	 & 	 105 $\pm$ 3	 & 159 \\ 
196	 & 	240.095830	 & 	-1.388480	 & 	8.652	 & 	36.374	 & 	1.0 $\pm$ 0.1	 & 	 83 $\pm$ 3	 & 137 \\ 
197	 & 	240.126800	 & 	-1.412810	 & 	8.650	 & 	36.335	 & 	0.8 $\pm$ 0.4	 & 	 97 $\pm$ 12	 & 151 \\ 
198	 & 	240.140400	 & 	-1.374960	 & 	8.698	 & 	36.346	 & 	1.4 $\pm$ 0.6	 & 	 94 $\pm$ 12	 & 149 \\ 
199	 & 	240.150360	 & 	-1.415330	 & 	8.665	 & 	36.314	 & 	1.2 $\pm$ 0.1	 & 	 105 $\pm$ 3	 & 159 \\ 
200	 & 	240.160870	 & 	-1.407540	 & 	8.680	 & 	36.310	 & 	0.5 $\pm$ 0.2	 & 	 80 $\pm$ 9	 & 134 \\ 
\multicolumn{8}{c}{}\\
201	 & 	240.216730	 & 	-1.160040	 & 	8.970	 & 	36.409	 & 	1.0 $\pm$ 0.2	 & 	 104 $\pm$ 5	 & 159 \\ 
202	 & 	240.276740	 & 	-1.219370	 & 	8.954	 & 	36.326	 & 	1.0 $\pm$ 0.1	 & 	 90 $\pm$ 3	 & 145 \\ 
203	 & 	240.299410	 & 	-1.210700	 & 	8.979	 & 	36.312	 & 	0.6 $\pm$ 0.2	 & 	 90 $\pm$ 9	 & 144 \\ 
204	 & 	240.306750	 & 	-1.562050	 & 	8.630	 & 	36.101	 & 	1.0 $\pm$ 0.1	 & 	 84 $\pm$ 3	 & 139 \\ 
205	 & 	240.361430	 & 	-1.589380	 & 	8.642	 & 	36.041	 & 	1.0 $\pm$ 0.1	 & 	 89 $\pm$ 3	 & 143 \\ 
206	 & 	240.370770	 & 	-1.520710	 & 	8.718	 & 	36.074	 & 	1.3 $\pm$ 0.6	 & 	 100 $\pm$ 12	 & 155 \\ 
207	 & 	240.542190	 & 	-1.564570	 & 	8.797	 & 	35.909	 & 	0.8 $\pm$ 0.1	 & 	 86 $\pm$ 3	 & 141 \\ 
208	 & 	240.544860	 & 	-1.605910	 & 	8.757	 & 	35.883	 & 	1.4 $\pm$ 0.4	 & 	 78 $\pm$ 7	 & 132 \\ 
209	 & 	240.547530	 & 	-1.485240	 & 	8.881	 & 	35.950	 & 	1.4 $\pm$ 0.1	 & 	 75 $\pm$ 2	 & 130 \\ 
210	 & 	240.580210	 & 	-1.563240	 & 	8.826	 & 	35.878	 & 	1.3 $\pm$ 0.3	 & 	 77 $\pm$ 6	 & 131 \\ 
\multicolumn{8}{c}{}\\
211	 & 	240.581540	 & 	-1.608570	 & 	8.781	 & 	35.851	 & 	0.8 $\pm$ 0.2	 & 	 87 $\pm$ 6	 & 141 \\ 
212	 & 	240.986681	 & 	0.436293	 & 	11.148	 & 	36.691	 & 	0.4 $\pm$ 0.2	 & 	 91 $\pm$ 12	 & 147 \\ 
213	 & 	241.059251	 & 	0.027519	 & 	10.779	 & 	36.401	 & 	0.8 $\pm$ 0.2	 & 	 88 $\pm$ 7	 & 144 \\ 
214	 & 	241.070099	 & 	0.439281	 & 	11.210	 & 	36.624	 & 	1.0 $\pm$ 0.4	 & 	 70 $\pm$ 9	 & 126 \\ 
215	 & 	241.071945	 & 	0.418463	 & 	11.190	 & 	36.611	 & 	0.7 $\pm$ 0.1	 & 	 71 $\pm$ 4	 & 127 \\ 
216	 & 	241.134037	 & 	-0.022007	 & 	10.781	 & 	36.311	 & 	0.3 $\pm$ 0.1	 & 	 107 $\pm$ 7	 & 163 \\ 
217	 & 	241.154955	 & 	0.004965	 & 	10.823	 & 	36.309	 & 	0.4 $\pm$ 0.2	 & 	 99 $\pm$ 8	 & 155 \\ 
218	 & 	241.159661	 & 	0.053231	 & 	10.876	 & 	36.332	 & 	0.7 $\pm$ 0.1	 & 	 67 $\pm$ 5	 & 123 \\ 
219	 & 	241.167785	 & 	0.699604	 & 	11.547	 & 	36.689	 & 	0.9 $\pm$ 0.1	 & 	 74 $\pm$ 4	 & 130 \\ 
220	 & 	241.181238	 & 	0.731057	 & 	11.589	 & 	36.696	 & 	0.8 $\pm$ 0.2	 & 	 67 $\pm$ 6	 & 123 \\ 
\multicolumn{8}{c}{}\\
221	 & 	241.214128	 & 	0.740654	 & 	11.622	 & 	36.674	 & 	1.0 $\pm$ 0.1	 & 	 64 $\pm$ 2	 & 120 \\ 
222	 & 	241.312445	 & 	-0.104714	 & 	10.822	 & 	36.117	 & 	0.5 $\pm$ 0.1	 & 	 96 $\pm$ 6	 & 151 \\ 
223	 & 	241.448553	 & 	0.805944	 & 	11.852	 & 	36.516	 & 	1.4 $\pm$ 0.3	 & 	 75 $\pm$ 6	 & 131 \\ 
224	 & 	241.451074	 & 	0.571298	 & 	11.612	 & 	36.383	 & 	1.7 $\pm$ 0.8	 & 	 64 $\pm$ 12	 & 120 \\ 
225	 & 	241.465961	 & 	0.588172	 & 	11.639	 & 	36.380	 & 	0.8 $\pm$ 0.1	 & 	 60 $\pm$ 4	 & 116 \\ 
226	 & 	241.488214	 & 	0.590887	 & 	11.658	 & 	36.363	 & 	1.2 $\pm$ 0.6	 & 	 72 $\pm$ 13	 & 128 \\ 
227	 & 	241.504434	 & 	0.825280	 & 	11.911	 & 	36.480	 & 	2.1 $\pm$ 0.1	 & 	 86 $\pm$ 1	 & 142 \\ 
228	 & 	241.524608	 & 	0.780546	 & 	11.878	 & 	36.438	 & 	0.9 $\pm$ 0.4	 & 	 78 $\pm$ 11	 & 134 \\ 
229	 & 	241.683613	 & 	0.521306	 & 	11.721	 & 	36.162	 & 	0.3 $\pm$ 0.1	 & 	 87 $\pm$ 9	 & 143 \\ 
230	 & 	241.722132	 & 	0.102803	 & 	11.319	 & 	35.896	 & 	0.6 $\pm$ 0.2	 & 	 76 $\pm$ 7	 & 132 \\ 
\multicolumn{8}{c}{}\\
231	 & 	241.732762	 & 	0.013001	 & 	11.235	 & 	35.836	 & 	0.6 $\pm$ 0.1	 & 	 75 $\pm$ 3	 & 131 \\ 
232	 & 	241.754781	 & 	0.320812	 & 	11.565	 & 	35.991	 & 	0.9 $\pm$ 0.1	 & 	 62 $\pm$ 4	 & 118 \\ 
233	 & 	241.756613	 & 	0.104295	 & 	11.345	 & 	35.868	 & 	1.5 $\pm$ 0.3	 & 	 85 $\pm$ 5	 & 141 \\ 
234	 & 	241.767043	 & 	0.093198	 & 	11.341	 & 	35.853	 & 	0.5 $\pm$ 0.1	 & 	 81 $\pm$ 3	 & 137 \\ \hline
\end{tabular}
Note: Column 1 shows the star id; column 2 represents right ascension ($\alpha$) in increasing order; column 3 denotes the declination ($\delta$); columns 4 \& 5 show the Galactic longitude and latitude, respectively; column 6 represents the values of degree of polarization with errors; column 7 represents the polarization position angles with errors measured from the North increasing towards the East, and column 8 denotes the Galactic polarization position angles measured from the Galactic North.
\end{threeparttable}
\end{center} 
\end{table*}

\bsp	
\label{lastpage}
\end{document}